\documentclass[journal]{IEEEtran}

\usepackage{cite}

\usepackage[ruled,vlined]{algorithm2e}
\usepackage{cite}
\usepackage{amsmath,amssymb,amsfonts}
\usepackage{algorithmic}
\usepackage{graphicx}
\usepackage{textcomp}
\usepackage{xcolor}
\include{pythonlisting}
\usepackage{amsthm}
\usepackage{comment}
\usepackage{enumerate}
\usepackage{caption}
\usepackage{subcaption}
\usepackage[ruled,vlined]{algorithm2e}
\usepackage{wrapfig}
\usepackage{titlesec}
\usepackage{lipsum}% just to generate text for the example

\titlespacing*{\section}
{0pt}{2mm}{2mm}

\titlespacing*{\subsection}
{0pt}{1mm}{1mm}

\begin{document}

\font\myfont=cmr12 at 20pt
\title{{\myfont Contention Based Proportional Fairness (CBPF) Transmission Scheme for Time Slotted Channel Hopping Networks}}
\author{Bommisetty Lokesh and T G Venkatesh
\thanks {Lokesh Bommmisetty and T G Venkatesh are with the Department of Electrical Engineering, Indian Institute of Technology Madras, Chennai- 600036, INDIA. (e-mails:lokesh.jun12@gmail.com, tgvenky@ee.iitm.ac.in)}}
%\thanks{$\dagger$Corresponding author}
\date{}

%\author{Bommisetty Lokesh and T G Venkatesh}

%\markboth{}
%{Shell \MakeLowercase{\textit{et al.}}: Bare Demo of IEEEtran.cls for Journals}

\maketitle

\begin{abstract}
% In this letter, we carry out the performance analysis of multichannel slotted ALOHA based scheduled network under saturated traffic condition where the scheduling algorithms decide the transmission probabilities of the network nodes in any slot based on the priority order. First, we derive the performance parameters like throughput, delay and energy spent in an unscheduled network where all the network nodes transmit in a slot with same probability. Then, the performance of the scheduled network is analysed and plots are obtained for the said performance parameters. The analytical results are validated by substantial simulations. 
%The emergence of the Internet of Things(IoT) allowed the number of connected devices and the amount of transmitted data over the internet to increase enormously. Recent Medium Access Control (MAC) standards for IoT networks such as Time Slotted Channel Hopping (TSCH) are modelled as multichannel slotted ALOHA network. The demands of different devices connected over the internet are different due to the diverse applications. Hence the nodes are scheduled to access the channel to achieve the requirements of applications. In this context, this letter evaluates the performance metrics such as throughput, delay and energy spent per packet transmission for a scheduled multichannel slotted ALOHA network. Substantial simulations validate the analytical results obtained. It is shown that the network can achieve optimal performance even in the case of scheduling.     
\textcolor{teal}{Time Slotted Channel Hopping (TSCH) is a Medium Access Control (MAC) protocol introduced in IEEE802.15.4e standard, addressing low power requirements of the Internet of Things (IoT) and Low Power Lossy Networks (LLNs).} The 6TiSCH Operation sublayer (6top) of IEEE802.15.4e defines the schedule that includes sleep, transmit and receive routines of the nodes. However, the design of schedule is not specified by the standard. In this paper, we propose a contention based proportional fairness (CBPF) transmission scheme for TSCH networks to maximize the system throughput addressing fair allocation of resources to the nodes. We propose a convex programming based method to achieve the fairness and throughput objectives. We model TSCH MAC as a multichannel slotted aloha and analyse it for a schedule given by the 6top layer. Performance metrics like throughput, delay and energy spent per successful transmission are derived and validated through simulations. \textcolor{red}{The proposed CBPF transmission scheme has been implemented in the IoT-LAB public testbed to evaluate its performance and to compare with the existing scheduling algorithms. }
\end{abstract}
\section{Introduction}\label{sec:introduction}

%\IEEEPARstart{I}{n} recent times, the things are evolving at a rapid pace and are getting connected to internet. This evolution came to an extent where these things are able to communicate among themselves and form a network called Internet of Things(IoT). Wireless sensor and actuator networks(WSANs) are perceived as a dominating technology in many application domains and are expected to play a crucial role in the realization of the future of IoT. WSANs were already proved effective in many application domains such as smart cities, healthcare, industrial automation etc.,\cite{hashemi2014intra,tsai2007zigbee}.
 
\IEEEPARstart{I}{n} recent times, deployment of Low Power and Lossy Networks(LLN) has been boosted by the emergence of Internet of Things (IoT)\textcolor{blue}{\cite{hashemi2014intra,elsts2020empirical,Guglielmo2016IEEE8}}. LLNs consists of low complexity resource-constrained devices with constraints on processing power, memory, energy as they are mostly battery operated. Wireless sensor and actuator networks(WSANs) are perceived as a dominating technology in many application domains and are expected to play a crucial role in the realization of the future of IoT. WSANs were already proved effective in many application domains such as smart cities, healthcare, industrial automation etc., which are a result of rapidly expanding IoT networks \textcolor{blue}{\cite{hashemi2014intra,elsts2020empirical}}. Traffic generated by such devices will be sporadic but demand low latency and energy efficiency because of the devices being battery operated.

The IEEE 802.15.4e standard based on Time Slotted Channel Hopping (TSCH), addresses the low energy requirements of IoT networks. TSCH behavioral mode of IEEE 802.15.4e combines time slotting and multichannel communication with channel hopping capabilities to provide predictable latency, communication reliability and energy efficiency\cite{Guglielmo2016IEEE8}. The channel hopping mechanism of TSCH allows overcoming the impact of multipath fading by changing the communication channel for every transmission. TSCH is widely implemented in several use cases like industrial monitoring and automation, smart homes, healthcare and environment monitoring \cite{elsts2020empirical}. TSCH devices transmit in resource units called cells which are defined by the timeslot index and the channel offset. 

\begin{figure}
    \centering
    \includegraphics[scale=0.4]{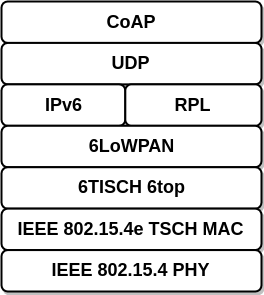}
    \vspace{2mm}
    \caption{IEEE 802.15.4e protocol stack\cite{thubert2015architecture}}
    \label{fig:protocol stack}
    
\end{figure}
A sublayer called 6TiSCH operation sublayer (6top) is built on top of 802.15.4e TSCH MAC as shown in Fig. \ref{fig:protocol stack} to make the routing protocol for LLNs (RPL) compatible with 802.15.4e devices. RPL is the routing protocol developed to address several application specific optimizing objectives like minimizing energy, minimizing latency in LLNs \cite{winter2012rpl}. 6top layer also enables the adoption of IPv6 over TSCH\cite{thubert2015architecture}. Apart from these functionalities, 6top sublayer additionally enables scheduling to provide differential treatment to network devices, hereafter termed as nodes, as the demands at each node is different to satisfy the application demands such as low end-to-end latency, high reliability and energy efficiency \cite{9203871}. 

The 6top layer in IEEE 802.15.4e TSCH will only define the execution of a given schedule without specifying how to create and maintain a proper link scheduling. TSCH schedule defines the share of resources a node can use which is determined by the scheduling algorithms. In TSCH, scheduling helps to effectively allocate resource cells to wireless links. The design of a schedule will have direct impact on the network efficiency, including node energy consumption, latency, network throughput and reliability. These scheduling algorithms may  be targeted to optimize different performance parameters constrained over the load conditions, application demand, latency requirement etc., at each node.

Several TSCH scheduling algorithms have been proposed in literature. They can be broadly classified into three categories: centralized, distributed and autonomous scheduling \cite{kharb2019survey}. In centralized scheduling, a central entity or controller builds the schedule and broadcasts it to all the nodes in network. In distributed scheduling, the schedule of a node is built on the basis of its negotiations with the neighbouring nodes. In the third type of scheduling called autonomous scheduling, the schedule of each node is constructed by that node autonomously. These three types of scheduling are in the order of reduction of network infrastructural costs. 

In this paper, our attention is on the distributed scheduling algorithms. Several distributed algorithms proposed in literature are contention free in nature. Generally, contention free schedules require a well built and centralised infrastructure for the network which is not cost effective \cite{8481452}. Contention free schedules also induce large overhead messages to the payload data. The IoT and sensor networks in general generate sporadic traffic in which case reservation based scheduling algorithms are highly inefficient in terms of resource utilisation. Besides, in dense networks reservation based schedules introduce long transmission delays. Hence contention free scheduling algorithms become highly inefficient in dense IoT networks. Whereas the contention based scheduling achieves higher link throughput for sporadic traffic however inducing the frequent collisions resulting in retransmission of data packets.

\textcolor{red}{Medium access algorithms are used in wireless networks to control access to a shared wireless medium, and thereby reduce collisions, ensure high system throughput, and distribute the available bandwidth fairly among the competing streams of traffic. The problem of fair rate control at the transport layer have been extensively researched, however the fair resource allocation at the MAC design has not been adequately addressed in TSCH networks\textcolor{blue}{\cite{8481452}}.}

Aim of this paper is to propose a contention based distributed scheduling algorithm for TSCH networks that achieves proportional fairness. We model the TSCH MAC layer as a multichannel slotted aloha system where the transmission probability of each node in a slot is given by the scheduling algorithm. Because of the scheduling, the network nodes behave heterogeneously while contending for the shared resources. In view of growing IoT networks and application diversity, it is of extreme importance that the performance of a TSCH network under the heterogeneous behaviour to be analysed.  

\textcolor{red}{
The major contributions of our paper are as follows.
\begin{itemize}
    \item We propose a proportional fairness contention based schedule for TSCH networks.
    \item We model and analyze the TSCH MAC protocol as a multichannel slotted aloha wherein the transmission probability of a node in a time slot is given by the schedule.
    \item We mathematically derive the performance metrics, namely throughput, packet transmission delay and energy spent per successful transmission of TSCH network for a given schedule.
    \item We have implemented and evaluated the proposed CBPF MAC protocol in the IoT-LAB \textcolor{blue}{\cite{adjih2015fit}} public testbed and compared its performance with the existing scheduling algorithms.
\end{itemize}
}
\textcolor{red}{
The reminder of the paper is organised as follows. We discuss the background and literature of TSCH scheduling algorithms to form the motivation of our work in Section \ref{sec:Related Work}. The IEEE 802.15.4 e TSCH network is  modelled as a generic multichannel slotted aloha network. We formally define the possible conflicts during transmissions in our system model mathematically in Section \ref{system model}. We formulate the problem of finding optimal transmission scheme for the random access in TSCH networks as a convex optimisation problem in Section \ref{sec:algorithm}. In Section \ref{Analysis}, we apply the random access transmission scheme proposed in Section \ref{sec:algorithm} to determine the transmission probabilities of the nodes in a data collector network. The performance of the proposed transmission scheme on the data collection network is mathematically analysed in Section \ref{performance}. We present the analytical results and validate them using simulations and testbed implementation along with the comparative analysis with the existing scheduling algorithms in Section \ref{results}. Finally we conclude our paper in Section \ref{conclusion}.}

\section{Background and Related Work} \label{sec:Related Work}
TSCH schedule can be explained in two steps. Firstly, a scheduling algorithm (also known as scheduling function SF) specifies the design of schedule to the 6top sublayer. Scheduling algorithm decides the number of cells to be added/deleted from the schedule of a node. However, the implementation of a schedule is done in 6top layer by the 6top protocol (6P) which is defined by IETF in the standard.  

\subsection{6top Protocol (6P)}
The 6top protocol (6P) is defined in RFC 8480 \cite{wang20186tisch}. 6P transaction is a pairwise negotiation mechanism for cell negotiation thus allowing the neighbours to agree on the schedule in distributed manner. A 6P transaction can be an ADD/Delete/Relocate request of cells to/from/in a schedule between two nodes. 6P transaction can consist of two or three steps. To schedule a link from node A to node B, in 2-step 6P transaction, node A proposes the set of cells to be allocated. While in 3-step 6P transaction, node B puts forward the cells to be allocated. The scheduling algorithm must specify whether to use 2-step or 3-step transaction. In either type of transactions, once a node proposes a set of cells for the schedule, it locks the cells and waits for the confirmation of the final list of cells from its neighbour. Once the 6P transaction is complete, the cells which are agreed upon will be added/deleted to/from the schedule of link from node A to node B.
\subsection{Scheduling algorithms}
In their survey \cite{kharb2019survey}, Kharb \textit{et al.} presented more than forty TSCH scheduling algorithms and classified them into centralized, distributed, autonomous and hybrid scheduling algorithms. Distributed scheduling algorithms are well suited for the networks with dynamic topologies and also for large and dense networks. Hence they occupy the major share (55\%) of the total scheduling algorithms. Most of these scheduling algorithms consider the required bandwidth, system throughput, energy spent per packet transmission etc., as the optimizing parameters in their design. 

The IETF 6TiSCH adopts a default scheduling algorithm called minimal scheduling function (MSF) in RFC8180. MSF uses a combination of Id-based scheduling and neighbor to neighbor scheduling which means a hybrid of autonomous and distributed scheduling. Palettella \textit{et al.} proposed a distributed scheduling scheme called On-the fly (OTF)\cite{palattella2015fly}. OTF calculates the number of cells to be allocated to each link based on the estimated traffic on that link. But as the links are not aware of the traffic on the neighbouring links, OTF is prone to collisions and exhibits a poor performance for dense networks.
Kravelska \textit{et al.} proposed a queue length based load balancing distributed scheduling algorithm called local voting \cite{vergados2017toward}. In local voting, the authors considered a weighted allocation of cells to a link depending on their neighbouring link's traffic. 

A static TSCH scheduling algorithm proposed by Park \textit{et al.} reduces the control message collisions to achieve high packet delivery ratio \cite{8576990}. This is achieved by individually defining the broadcast and unicast slots. Kim \textit{et al.} proposed a link based scheduling algorithm called ALICE considering the possible collisions into the design of scheduling \cite{kim2019alice}. \textcolor{teal}{ALICE does a mapping between link and their transmission cells using a hash table, which becomes NP hard to find the optimal mapping for larger and time-varying networks. To handle the unpredictable changes in the network due to time varying traffic and network topology, Jeong \textit{et al.} proposed a traffic aware scheduling algorithm called TESLA \cite{TESLA}. TESLA allows each to change the slotframe size according to its traffic load. However, this slotframe size is propagated to the neighbouring nodes. In OST \cite{jeong2020ost}, depending on the traffic load, nodes dynamically adapt the frequency of transmitting cells, minimising the energy consumption.} 
%Orchestra \cite{duquennoy2015orchestra} is proposed by Duquennoy \textit{et. al} to maximize the reliability based on a collision free schedule. 
%Orchestra has a policy of assigning unique slot to all nodes in the network. For this to be possible, the periodicity of a node's transmission is longer than the number of nodes in the network. Even though the reliability is maximized, the latency of the network is large, thus making Orchestra unsuitable for large scale networks or high traffic applications. 

Javan \textit{et al.} proposed a combinatorial multi armed bandit approach to achieve optimal throughput in TSCH networks \cite{CMAB}. In their model, the candidate cells for scheduling are determined using bipartiate graphs and coloring methods in graph theory based on the load and network topology. This approach completely avoids the collision, but at higher loads, the greedy approach of cell allocation may limit resources to the nodes with lower demand resulting in an unfair resource allocation. 

All these scheduling algorithms calculate the number of cells to be added/deleted from a particular node's schedule. Once the scheduling algorithm decides the number of cells to be added/deleted, 6top protocol (6P) triggers a negotiation between neighbouring nodes to decide on the location of cells to be added/deleted in each node's schedule. \textcolor{red}{Besides, these scheduling algorithms \textcolor{blue}{\cite{palattella2015fly,vergados2017toward,kim2019alice,8576990,CMAB}} quantitatively does not measure the fair resource allocation. Ojo \textit{et al.} emphasised the need for fair scheduling algorithms and commented on the incompatibility of the general wireless network scheduling algorithms to TSCH and IIoT networks. The authors proposed a heuristic min-max fair scheduling algorithm in their work \textcolor{blue}{\cite{8481452}}.  }

 \subsection{Motivation}
 The 6P transactions occur simultaneously at all the nodes in network. Hence there is a chance that two interfering nodes (need not share a schedule) may put forward the same resource units as the candidate cells which are picked in random from the available set of cells inherently leading to contention. Hence negotiation procedure increases the convergence delay due to the collision prone scheduled cells. This phenomenon gets pronounced as the network grows large or becomes denser. Even worse, this approach is inefficient for dynamic network conditions as the renegotiation is time consuming and implies packet drops before the convergence of new schedule \cite{9096372}. This inevitably lowers the throughput of the network. Besides, the candidate cells get locked for the entire duration of 6P transaction, irrespective of it being a successful or failure, reducing the utilization of resources. Scheduling algorithms in literature do not take into the account the possible contention during 6P transactions. Hence they are not designed to minimize the collisions during 6P transactions. \textcolor{teal}{But our algorithm takes into account the load of the network at each node and fixes the transmission probability of each node so as to maximise the fairness throughput. The transmission probabilities solved using our CBPF algorithm can be used to construct a slotframe instead of doing it in pseudo-random method.} \textcolor{red}{Proportional fairness was first introduced by Frank Kelly in \textcolor{blue}{\cite{kelly1997charging}} based on changing rate control for elastic traffic in computer network services. Proportional fairness maximises the overall utility rate allocations with a logarithmic utility function. Moreover, in Aloha networks, the optimization problem for attaining proportional fairness is convex and separable \textcolor{blue}{\cite{massoulie2007structural}}, which allows us to develop computationally simple algorithms for arriving at the transmission probabilities that attain the optimal rates.}
 
 Unique features of our paper differentiating from the existing literature are as follows.
 \begin{itemize}
 \item  To the best of our knowledge, ours is the first work to propose a contention based transmission scheme that maximizes the proportional fairness throughput of the TSCH network. 
 \item Several works in literature\cite{7899546,8418137,alderisi2015simulative} have analysed the performance of TSCH network governed by a schedule only through simulations. In contrast, we mathematically analyse the performance of TSCH network governed by our proposed transmission scheme and validate the results using simulations. 
 \end{itemize}

\section{SYSTEM MODEL}
 \label{system model}
%We consider an IEEE802.15.4e-TSCH network consisting of $N$ nodes transmitting data to the border router over $M$ non-interfering channels. We consider a slotted transmission system, where the timeslot is long enough for a node to transmit its MAC frame and receive acknowledgement from the border router\cite{palattella2015using}. As demonstrated later in Section IV, a 6top schedule can be translated into transmission probabilities of nodes. Once a node chooses a slot to transmit, it selects a channel among $M$ available channels randomly. We assume the saturated traffic condition, where all the nodes have a packet to transmit at any time slot. This assumption lets us find the worst case performance of an IoT network. We consider a transmission to be successful only if no other node is transmitting in the same channel at same time slot. ise.

We model the IEEE 802.15.4e-TSCH as a generic multichannel slotted aloha network as follows. Our model considers a 6TiSCH network built by the Routing Protocol for Low-Power and Lossy Networks (RPL) \cite{winter2012rpl}. The network consists of a finite set $\mathcal{N}=\{1,2,3,\dots,N\}$ of nodes and operates in discrete time slots.  \textcolor{red}{Each node sends its data to the sink node over multiple hops. Hence a node's transmissions consists of data generated by itself and also the forwarding data of other nodes to the sink node}. Without loss of generality, we consider a single-sink model although the proposed algorithm works with multiple sinks. Let $\mathcal{D}_n\subseteq \mathcal{N}\setminus n$ denote the subset of relay nodes which can be used by node $n$ to send data to sink. A node $n$ may attempt transmission of a packet to one of the nodes $m \in \mathcal{D}_n$ in one of the channels chosen randomly in a time slot. We define a link $(n,m)$ to denote a directed communication from the transmitter node $n$ to the receiver node $m$, where $n\neq m$, in a specific time slot and a channel. Let $\mathcal{S}=\{(n,m)|n\in \mathcal{N},m \in \mathcal{D}_n\}$ denote the set of all links in a network. 

Time is divided into slots with slot index $t=0,1,2,\dots$. A time slot $t$ is long enough for the transmission of a data packet from source node to destination node and for the reception of corresponding acknowledgement. \textcolor{red}{The available bandwidth in IEEE 802.15.4e TSCH networks is divided into orthogonal channels.} Let the number of non interfering channels available to TSCH mechanism be denoted by $M$. The set of available channels is denoted by $\mathcal{M}=\{ch_i| 1\leq i \leq M\}$. A transmission occurs on a link using a resource unit called cell which is denoted by the time slot index and the channel offset as $(t,ch_i)$. The transmission on a link $(n,m)$ using the cell $(t,ch)$ is denoted by $c_{(n,m)}^{(t,ch)}$, where
\begin{equation}
 c_{(n,m)}^{(t,ch)}= \begin{cases}
 1, & \text{if transmission occurs on $(n,m)$ in cell $(t,ch)$}\\
 0, & \text{if no transmission on $(n,m)$ in cell $(t,ch)$}
 \end{cases}
 \label{c}
\end{equation}
\begin{comment}
We assume that a node cannot simultaneously transmit or receive on two or more different links. The set of all the links violating this assumption for a particular link $(n,m)$ is denoted and defined as 
\begin{equation}\mathcal{I}^{(a)}_{nm}=\{(l,m)|l\in \mathcal{D}_m\setminus n\}\lor \{(n,k)\mid k\in \mathcal{D}_n\setminus m\} \label{Ia} \end{equation}
We also assume that all the links between the nodes of network are half duplex. Hence, the simultaneous reception and transmission is not possible at a node. The set of links violating the half duplicity for a given link $(n,m)$ is defined as 
\begin{equation}
    \mathcal{I}^{(b)}_{nm}=\{(l,k)|(l=n,k\in \mathcal{D}_n\setminus m)\lor (k=m,m\in \mathcal{D}_l,l \neq n)\}
    \label{Ib}
\end{equation}
We call these two sets as primary conflicting links and denote it by $\mathcal{I}_{nm}^p=\mathcal{I}^{(a)}_{nm}\cup \mathcal{I}^{(b)}_{nm}$ and can be alternatively defined as
\end{comment}
\begin{figure}[t]
\begin{subfigure}{0.58\linewidth}
  \centering
  % include first image
  \includegraphics[width=\linewidth]{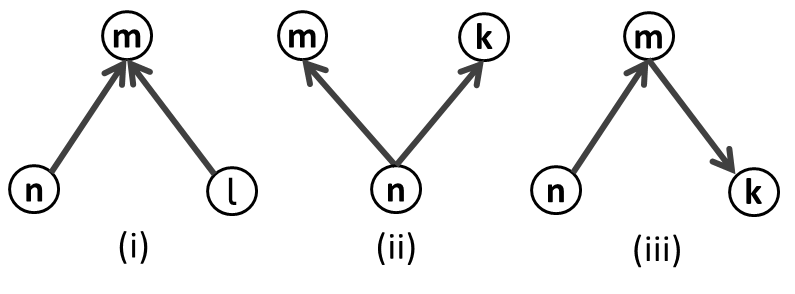}
     \subcaption{primary conflicting links}
     \label{fig:primaryconflict}
\end{subfigure}
\begin{subfigure}{.4\linewidth}
  \centering
  % include second image
  \includegraphics[width=\linewidth]{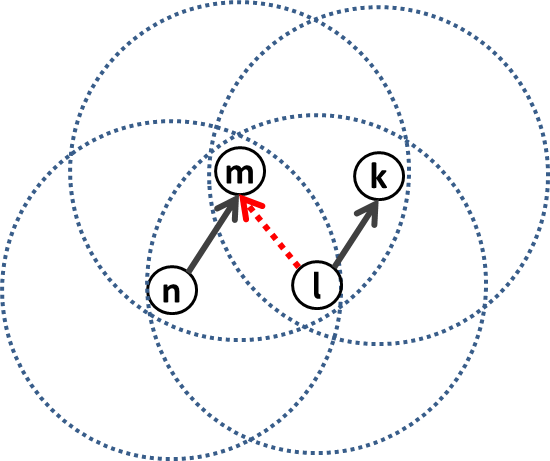}
    \subcaption{Secondary conflicting links}
     \label{fig:secondaryconflict}
\end{subfigure}
\caption{Examples of conflicting links in TSCH network}
\label{fig:fig}
\end{figure}
\subsection{Primary Conflict}
We consider the following reasonable assumptions on the communication capability of nodes in the network. Nodes do not have multi packet transmission / reception capability. In other words
\begin{enumerate}[(i)]
    \item A node cannot simultaneously transmit on two or more links and cannot simultaneously receive over two or more links. 
    \item A node cannot transmit and receive simultaneously.
\end{enumerate}
The above two conditions are called primary conflicting conditions. \textcolor{red}{Primary conflict refers to packet loss irrespective of the channel on which the communication occurs.} Fig \ref{fig:primaryconflict} illustrates different scenarios when primary conflict occurs. The set of links  which are in primary conflict with the transmission on the link $(n,m)$ is denoted and defined as follows
\begin{equation}
    \mathcal{I}_{nm}^p=\big\{(l,k)|\{l,k\}\cap \{n,m\}\neq\phi\big\}
    \label{I1}
\end{equation}
 A link $(l,k)$ such that $l=n$ or $m=k$ violate the condition (i). If $l=m$ or $k=n$, then the link $(l,k)$ violates the condition (ii). The transmissions on primary conflicting links are conditioned using equation \eqref{c} as 
 \begin{equation}
     c_{(n,m)}^{(t,ch_i)}c_{(l,k)}^{(t,ch_j)}=0, \text{if } (l,k)\in \mathcal{I}_{nm}^p
 \label{primeconf}
 \end{equation}

Equation \eqref{primeconf} indicates that transmission cannot occur on two primary conflicting links at the same time slot $t$, even on different channels $ch_i$ and $ch_j$.

\subsection{Secondary Conflict}
Secondary conflict occurs between the transmissions taking place on distinct links in the same inference range choosing the same channel in the same slot. \textcolor{red}{It stems from the fact that a receiver cannot decode an incoming packet in a channel $ch$ if another node in its neighborhood is also transmitting on the same channel $ch$ in the same time slot $t$ due to channel interference}. Let $\mathcal{N}_n$ be the set of nodes within the interference range of node $n$. It can be seen that $\mathcal{D}_n\subseteq \mathcal{N}_n$ (one hop transmission is possible only within the interference range). The set of links that interfere with the ongoing transmission on link $(n,m)$ is denoted as
\begin{equation}
    \mathcal{I}_{nm}^s=\{(l,k)|m\in \mathcal{N}_l \lor k\in \mathcal{N}_n \}
    \label{I2}
\end{equation}
\textcolor{red}{Equation \eqref{I2} gives the set of links which interfere with the transmissions on $(n,m)$ if operated on same channel and an example scenario is given in Fig. \ref{fig:secondaryconflict}.} It can be seen from \eqref{I1} and \eqref{I2} that $ \mathcal{I}_{nm}^p \subseteq \mathcal{I}_{nm}^s$. Also, note that the links in $ \mathcal{I}_{nm}^s\setminus \mathcal{I}_{nm}^p$ will not cause interference to link $(n,m)$ if operated on a different channel even in the same time slot. We call the links in $ \mathcal{I}_{nm}^s\setminus \mathcal{I}_{nm}^p$ as secondary conflicting links of $(n,m)$. The secondary conflicting links cannot transmit on the same time slot and the same channel which is formally written as follows.
 \begin{equation}
     c_{(n,m)}^{(t,ch)}c_{(l,k)}^{(t,ch)}=0, \text{if } (l,k)\in \mathcal{I}_{nm}^s \setminus \mathcal{I}_{nm}^p
 \label{secondconf}
 \end{equation}
 Equation \eqref{secondconf} indicates the interference constraint or the secondary conflict constraint. Fig. \ref{fig:secondaryconflict} shows a scenario of secondary conflict between two links. 
 
 Following the “Multi Channel Slotted Aloha"  random access
strategy, each node $n$ in TSCH network transmits its data or relays the forwarded data to the sink node in each time slot through a relay node $m\in \mathcal{D}_n$ with probability $\tau_{nm}$, \textcolor{red}{independent of other nodes and of the past transmission history.} When node $n$ does a transmission, it randomly chooses a particular channel $ch_i \in \mathcal{M}$ for transmission.

\begin{figure}
        \centering
        \includegraphics[width=\linewidth]{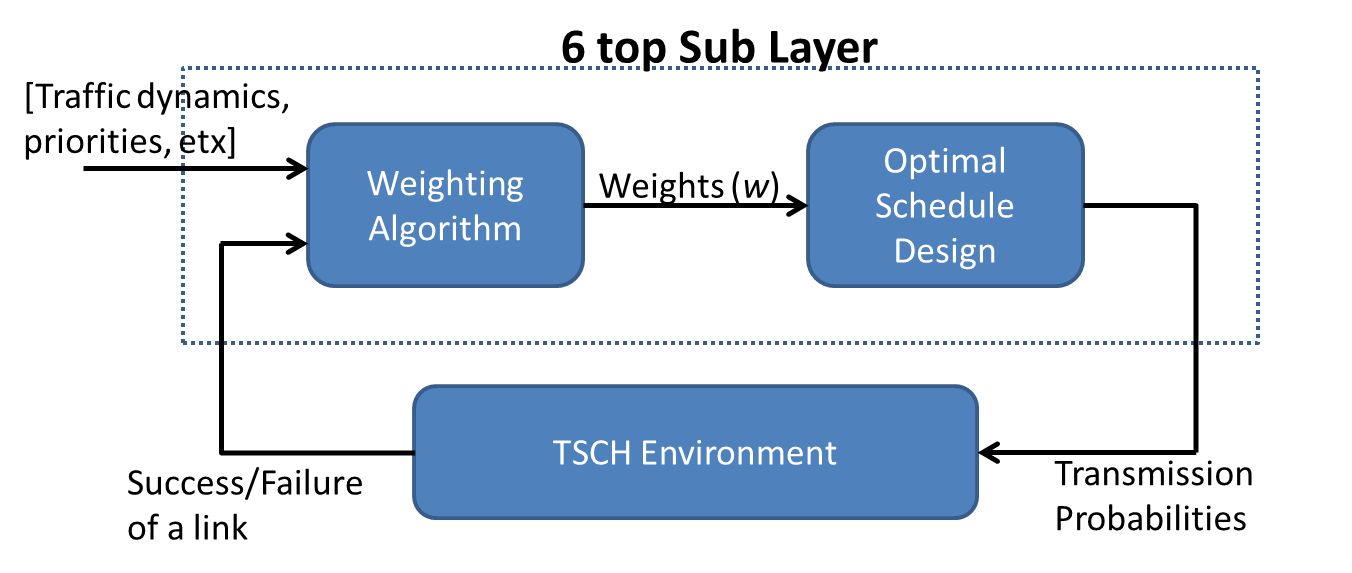}
        \caption{Block Diagram of the proposed MAC scheme}
        \label{fig:blockdia}
    \end{figure}

In nutshell, the methodology of our work to arrive at the optimal transmission scheme can be explained as follows. \textcolor{red}{Firstly, all links in the network are assigned weights according to the queue length i.e., the number of packets waiting to be transmitted on that particular link.} Transmissions in the network are restricted by two constraints. (i) No two primary conflicting links should be scheduled in the same time slot. (ii) The average number of transmitting nodes in a neighbourhood in the same time slot cannot exceed the number of available channels. With these two constraints, we determine the optimal transmission probabilities of all the links in $\mathcal{S}$ by maximising the weighted proportional fairness throughput utility function using the convex optimisation methods. Graphical illustration of the proposed random access algorithm in shown in Fig. \ref{fig:blockdia} using a block diagram. Once the optimal transmission scheme is determined, we mathematically analyse the performance of the TSCH network for a data collection network in terms of the link throughput and delay.
    
\section{Proposed Random Access Algorithm}\label{sec:algorithm}
In this section, we propose a contention based proportionally fair (CBPF) transmission scheme. \textcolor{red}{Proportional fairness strikes a balance between throughput maximisation and fair resource access among users in the network. The proposed scheduling algorithm gives the optimal transmission probability of each link in the network such that proportional fairness throughput is achieved in the network.} Following the random access mechanism of TSCH network outlined in Section \ref{system model}, the probability of the transmission over the link $(n,m)$ to be successful in a given time slot is given by
\begin{equation}
    \mu_{nm}=\tau_{nm}\prod_{(l,k)\in\mathcal{I}_{nm}^p}(1-\tau_{lk})\prod_{(l,k)\in\mathcal{I}_{nm}^s\setminus\mathcal{I}_{nm}^p}(1-\frac{\tau_{lk}}{M})
    \label{mu}
\end{equation}

Various terms in \eqref{mu} can be interpreted and explained as follows. $\tau_{nm}$ denotes the probability that a transmission occurs on link $(n,m)$. \textcolor{red}{The first product term ensures that there is no other transmissions on a link $(l,k)$ belonging to the primary conflicting zone of $(n.m)$, $\mathcal{I}_{nm}^p$. The second product term ensures that no link in the secondary conflicting zone of $(n,m)$ is transmitting on the same channel as that of the transmission over the link $(n,m)$.} Since there are $M$ channels the probability of link $(l,k)$ choosing the same channel as that of link $(n,m)$ is $\frac{1}{M}$.

Let us denote the vector of link success probabilities by $\boldsymbol{\mu}=(\mu_{nm}, (n,m)\in\mathcal{S})$, $\boldsymbol{\mu}\in [0, 1]^{S}$, where $S$ is the number of links in the network and is given by the cardinality of set $\mathcal{S}$, $S=|\mathcal{S}|$. From equation \eqref{mu}, it is clear that $\boldsymbol{\mu}$ is dependant on the set of the link transmission probability vectors $\boldsymbol{\tau}\in\mathcal{P}=\{(\tau_{nm},(n,m)\in \mathcal{S})\mid \tau_{nm}\in [0,1] \forall (n,m)\in\mathcal{S} \}$ and hence we denote the link success rate vector by $\boldsymbol{\mu}(\boldsymbol{\tau})$. We define $\mathcal{C}$, the capacity region of the network as the set of all possible link success probability vectors $\boldsymbol{\mu'}$, which can be majorised by vectors of the form $\boldsymbol{\mu}(\boldsymbol{\tau})$, and is represented as
\begin{equation}
    \mathcal{C}=\{\boldsymbol{\mu'}\in [0, 1]^S\mid \boldsymbol{\mu'}\preceq \boldsymbol{\mu}(\boldsymbol{\tau}) \text{ for some } \boldsymbol{\tau} \in \mathcal{P}\}
    \label{capacity region}
\end{equation}

 Let the packets arrive to each link $(n,m)$ according to a Poisson process with mean $\lambda_{nm}$. Let $\boldsymbol{\lambda} = (\lambda_{nm},(n,m)\in \mathcal{S})$, $\boldsymbol{\lambda} \in \mathbb{R}_+^S$. Each node maintains a counter $Q_{nm}$ for each of its outgoing links $(n,m)$, $m\in\mathcal{D}_n$ representing the queue length which is updated at the end of every slot as follows.
\begin{equation}
    Q_{nm}(t+1)=Q_{nm}(t)+a_{nm}(t)-h_{nm}(t)
    \label{Qupdate}
\end{equation}
where $Q_{nm}(0)=0$, $a_{nm}(t)$ is the number of new packets that arrived in time slot $t$ and $h_{nm}(t)=1$ if there was a successful transmission on link $(n,m)$ in time slot $t$ and $h_{nm}(t)=0$ otherwise. Intuitively, a good scheduling algorithm is expected to keep the number of packets waiting in the queues to be finite. To make this notion formal, we define that if $\boldsymbol{\lambda} \notin \boldsymbol{\Lambda}$ then no scheduling algorithm can keep queue lengths finite, where
\begin{equation}
    \boldsymbol{\Lambda} = \{\boldsymbol{\lambda} \in \mathbb{R}_+^S\mid \boldsymbol{\lambda} \leq \boldsymbol{\mu} \text{ component wise, for some } \boldsymbol{\mu}\in\mathcal{C}\}
    \label{valid lambda}
\end{equation}

\textcolor{red}{where $\boldsymbol{\Lambda}$ is the set of all arrival rate vectors such that the arrival rate at any node does not exceed it's service rate.} Each node assigns a weight to each outgoing link based on the number of packets queued for transmission on that link and updates it at the end of every slot. We denote the link weight by $w_{nm}$ for a link $(n,m)\in\mathcal{S}$. These link weights of each node $n$ are shared among the neighbouring nodes $\mathcal{N}_n$ for all $n\in \mathcal{N}$ periodically. The weight of each link is calculated as a function of queue length of that link as $w_{nm}=f(Q_{nm})$. Several earlier works proposed different classes of weight functions. We adapt the class of weight functions proposed by authors in \cite{9023554}, in particular we choose the weight function to be $f(x)=log(1+x)$, a positive, increasing and concave function of $x$. Hence the weight of a link $(n,m)$ in terms of its queue length $Q_{nm}$ is expressed as 
\begin{equation}
    w_{nm}=log(1+Q_{nm})
    \label{weights}
\end{equation}
Proportionally fair link success probabilities $\boldsymbol{\mu^*}$ are attained by the transmission probability vector $\boldsymbol{\tau^*}$ that \textcolor{red}{maximize the weighted proportional fairness objective function also known as utility function as given below}
\begin{equation}
    \begin{aligned}
     \boldsymbol{\tau^*}= \underset{\boldsymbol{\tau}\in \mathcal{P}}{argmax} \quad \sum_{(n,m)\in\mathcal{S}}w_{nm}\text{log}(\mu_{nm}(\boldsymbol{\tau}))
       % &=\sum_{(n,m)\in\mathcal{S}}w_{nm}\text{log}(\tau_{nm})+\sum_{(n,m)\in\mathcal{S}}w_{nm}(\Big{\sum_{(l,k)\in\mathcal{I}_{nm}^p}\text{log}(1-\tau_{lk})})\\ &\hspace{20}+\sum_{(n,m)\in\mathcal{S}}w_{nm}(\Big{\sum_{(l,k)\in\mathcal{I}_{nm}^s\setminus\mathcal{I}_{nm}^p}\text{log}(1-\frac{\tau_{lk}}{M})})
       \label{objective}
    \end{aligned}
    \end{equation}

\textit{Lemma 1:} The weighted proportional fairness objective function $F=\sum_{(n,m)\in\mathcal{S}}w_{nm}\text{log}(\mu_{nm}(\boldsymbol{\tau}))$ is concave with respect to $\boldsymbol{\tau}$.

\textit{Proof:} Proved in Appendix \ref{appendix}

\textcolor{red}{The maximum number of successful transmissions that can occur in a slot within the neighbourhood of a link is limited by the number of non-interfering channels $M$ or the number of competing nodes whichever is lower.} This gives us a constraint on the feasible link transmission probability vector. Hereby, we model the optimal link transmission probability problem as an optimization problem.
\begin{equation}
\begin{aligned}
\text{(P1):}\\
\underset{\boldsymbol{\tau}\in\mathcal{P}}{\text{minimize}}  &\quad -F=-\sum_{(n,m)\in\mathcal{S}}w_{nm}\text{log}(\mu_{nm})\quad\\
\text{subject to} &\quad \tau_{nm}+\sum_{(l,k)\in \mathcal{I}_{nm}^s} \tau_{lk}\leq M \quad\forall (n,m)\in \mathcal{S}
\label{P1}
\end{aligned}
\end{equation}
The objective function of the above optimization problem (P1) is convex since the function $F$ is concave according to \textit{Lemma 1}. The constraints are linear with $\boldsymbol{\tau}$. The domain $\mathcal{P}$ of the function $F$ is a hypercube in $\mathbb{R}^N$ and hence the domain of (P1) is also convex. Hence the problem (P1) is a convex optimization problem \cite{boyd2004convex}. It can be seen that the linear constraints always hold true for alteast one point in the domain $\boldsymbol{\tau} \in \mathcal{P}$. Hence the primal problem is said to be feasible.

As (P1) is a feasible convex optimisation problem, there exists a unique minimal point for the objective function and a unique link transmission probability vector $\boldsymbol{\tau}^*$. In the following section, we derive the closed form expression of the optimal transmission probability vector at the border router level of a TSCH network where it acts as a data collector.
\section{Optimal Transmission Probabilities for Data Collection Network}\label{Analysis}
In  this section, we mathematically derive the optimal transmission probabilities of nodes in the TSCH network at the border router. The border router receives data over multiple channels from multiple nodes forming a star topology. A network with this topology is often referred to as a data collection network. In this network setting, each node will have only one outgoing link and therefore the links can be identified by the transmitter node itself. The link access probabilities and the link success probabilities of the link connected to node $i\in \mathcal{N}$ are now denoted by $\tau_i$ and $\mu_i$ respectively. Further, all the links of the network are in interference conflict with each other. \textcolor{red}{Therefore, success probability of a link $i$ is  the product of $\tau_i$, probability of transmission on link $i$, and the probability that no other link $j$ transmits on the same channel as that of link $i$.}
\begin{equation}
    \mu_i = \tau_i \prod_{j=1, j\neq i}^N (1-\frac{\tau_j}{M}) 
    \label{mui}
\end{equation}
The convex optimization problem given by (P1) in \eqref{P1} can be adapted for the data collection network as
\begin{equation}
\begin{aligned}
\text{(P2):}\\
\underset{\tau \in \mathcal{P}}{\text{minimize}}  &\quad\mathcal{F}(\boldsymbol{\tau})=-\sum_{i=1}^Nw_{i}\text{log}(\mu_{i})\quad\\
\text{subject to} &\quad\quad \sum_{i=1}^N \tau_i \leq M
\label{P2}
\end{aligned}
\end{equation}
\textcolor{red}{where the condition that the optimization problem is subjected to refers to the average number of transmissions not exceeding the number of channels}. The convexity of the objective function has been already proved for the general case in Appendix \ref{appendix}. The objective function $\mathcal{F}(\tau):\mathbb{R}_+^N \rightarrow \mathbb{R}$ is also called as primal function. The constraint in \eqref{P2} is clearly a linear function of $\tau$, with the domain of $\tau$ being a hypercube in $\mathbb{R}_+^N$. Therefore the above problem (P2) is clearly a convex optimisation problem. Let the primal point $f^*$ of $\mathcal{F}(\boldsymbol{\tau})$ is defined as follows \textcolor{red}{\cite{boyd2004convex}}
\begin{equation}
    f^* = \underset{\tau \in \mathcal{P}}{min}\quad \mathcal{F}(\boldsymbol{\tau})
    \label{primalpoint}
\end{equation}
Lagrangian function \textcolor{red}{\cite{boyd2004convex}} for the convex optimization problem (P2), $\mathcal{L}:\mathbb{R}_+^N\times\mathbb{R}_+\rightarrow\mathbb{R}$ is given as
\begin{equation}
\begin{aligned}
    \mathcal{L}(\boldsymbol{\tau},\gamma)=\sum_{i=1}^N -w_i\log(\mu_i)+\gamma\Big(\sum_{i=1}^N\tau_i-M\Big)
    \end{aligned}
    \label{L2}
\end{equation}
where $\gamma \in \mathbb{R}_+$ is the Lagrangian multiplier. The dual function $\mathcal{G}$ of the convex problem (P2) and the dual point $g^*$ are given as follows \textcolor{red}{\cite{boyd2004convex}}
\begin{equation}
    \mathcal{G}(\gamma)=\underset{\boldsymbol{\tau} \in \mathcal{P}}{inf}\quad\mathcal{L}(\boldsymbol{\tau},\gamma)
    \label{dual2}
\end{equation}
\begin{equation}
    g^*=\underset{\gamma\geq 0}{max}\quad \mathcal{G}(\gamma)
    \label{dualpoint}
\end{equation}
where $\mathcal{P}$ is the domain of $\boldsymbol{\tau}$. Since the constraint in \eqref{P2} is feasible in $\mathcal{P}$, ($\exists \boldsymbol{\tau}^{'} \in \mathcal{P}: \sum_{i=1}^N\tau_i^{'}\leq M$), (P2) is a feasible convex optimisation problem for which the strong duality holds ($f^*=g^*$) by Slater's condition \cite{boyd2004convex}. Hence the Karush-Kuhn-Tucker (KKT) conditions \cite{boyd2004convex} can be used to find the optimal solution $(\boldsymbol{\tau}^*,\gamma^*)$. The KKT conditions for the above optimisation problem (P2) are given by
\begin{enumerate}
    \item $\sum_{i=1}^N \tau^*_i-M \leq 0$
    \item $\gamma^*(\sum_{i=1}^N \tau^*_i-M )=0$
    \item $\frac{\partial \mathcal{L}}{\partial \tau^*_i}=0$
\end{enumerate}
Here the first condition shows the feasibility of (P2). The second condition is called the complementary slackness condition. It says that if a dual variable is greater than zero (slack) then the corresponding primal constraint must be an equality (tight.) It also says that if the primal constraint is slack then the corresponding dual variable is tight (or zero). To find the optimal transmission probabilities, we apply condition 3 on \eqref{L2} as follows,
\begin{equation}
    \frac{\partial \mathcal{L}}{\partial \tau^*_i}=\frac{-w_i}{\tau^*_i}+\frac{\sum_{j=1, j\neq i}^N w_j}{M-\tau^*_i}+\gamma=0
    \label{diff2}
\end{equation}
 Rearranging \eqref{diff2}, we get a quadratic equation as follows.
\begin{equation}
    \gamma \tau_i^{*2}-(\gamma M-\sum_{j=1}^N w_j)\tau_i^*+w_iM=0 
    \label{quadratic}
\end{equation}
This quadratic equation has two roots of the form $\tau_i(\gamma)$ out of which only one of them will lie in the domain of (P2). \textcolor{red}{Since the problem (P2) is convex and is feasible in its domain, the uniqueness of the optimal point is expected.} The unique optimal solution $\boldsymbol{\tau}^*$ of (P2) is determined by finding the optimal dual variable $\gamma^*$ from \eqref{dualpoint} using first order optimisation methods like gradient descent or numerical methods. This concludes our discussion on the procedure to obtain the optimal transmission probability vector $\boldsymbol{\tau^*}$ which can be used to achieve the proportional fairness scheduling in TSCH network. 
\section{Performance Analysis}
\label{performance}
In this section, we analyse the performance of TSCH network operating using contention based proportional fairness scheduling algorithm  derived in Section \ref{Analysis}. In particular, we derive the throughput, delay and energy spent per successful packet transmission by a node in the data collection network. 
\subsection{Number of nodes transmitting in a slot}
    Let $\tau_i$, $i=1,2,\dots,N $ being the transmission probabilities of nodes in a data collection network. Let the random variable $K$ denote the number of nodes transmitting in a slot. $K$ can be treated as the sum of $N$ independent and non-identical Bernoulli random variables $I_j$. Where $I_j$'s indicate whether node $j$ transmits in a given time slot or not. 
    \begin{equation}
        K=\sum_{j=1}^N I_j\\
        \label{K}
        \end{equation}
        where
        \begin{equation}
  I_j =
    \begin{cases}
      1 & \text{w.p. $\tau_j$}\\
      0 & \text{w.p. 1-$\tau_j$}
    \end{cases} 
    \label{Ij}
    \end{equation}
\textcolor{red}{Equation \eqref{Ij} indicates that node $j$ transmits in a slot with probability $\tau_j$.} Random variable $K$ would follow a binomial distribution $B(k,N,\tau)$ if the indicator random variables were identical i.e., $\tau_i=\tau_j=\tau$. Since the indicator random variables are non identical, we determine the distribution of $K$ using characteristic function approach as follows. The characteristic function of $I_j$ is given as $\varphi_{I_j}(t)=1-\tau_j +\tau_j.exp(it)$ where $i=\sqrt{-1}$ \cite{ross1996stochastic}. Hence the characteristic function of $K$ is  
    \begin{equation}
    \begin{aligned}
        \varphi_K(t)=&\mathbb{E}[exp(itK)]=&\sum_{k=0}^NP_t(k;N).e^{itk}\\
         \end{aligned}
        \label{charecfunc}
    \end{equation}  
        where $P_t(k;N)$ is the probability that out of $N$ nodes, $k$ nodes transmit and $N-k$ nodes differ transmission in a given time slot. \textcolor{red}{Since $K$ is the sum of $N$ Bernoulli random variables, $\varphi_K(t)$ can be expressed as the product of individual characteristic functions $\varphi_{I_j}(t)$. Hence, \eqref{charecfunc} can be alternatively expressed from \eqref{K} and $\varphi_{I_j}(t)$ as follows}
        \begin{equation}
    \begin{aligned}
       \varphi_K(t)=&\prod_{j=1}^N\varphi_{I_j}(t) =&\prod_{j=1}^N(1-\tau_j+\tau_j.exp(it))
        \end{aligned}
        \label{charecfunc2}
    \end{equation}  
      Let us denote the IDFT of the $N+1$ length sequence $P_t(k;N)$ as $D(l)$. It can be seen that $D(l)$ can be obtained by substituting $t=\omega l$ in $\varphi_{K}(t)$ in \eqref{charecfunc} where $\omega = \frac{2\pi}{N+1}$. $D(l)$ which can be obtained from \eqref{charecfunc} and \eqref{charecfunc2} as follows\cite{ross1996stochastic}, \begin{equation}
        D(l)=\varphi_K(t)\mid_{t=\omega l}=\prod_{j=1}^N(1-\tau_j +\tau_j.exp(i\omega l))
        \label{IDFT}
    \end{equation}
    \textcolor{red}{Since $D(l)$ is the IDFT of the sequence $P_t(k;N)$, DFT of $D(l)$ will result in $P_t(k;N)$ which is the probability that $k$ out of $N$ nodes transmit.} Hence $P_t(k;N)$ is given by
    \begin{equation}
    \begin{aligned}
        P_t(k;N)&=\frac{1}{N+1}\sum_{l=0}^N D(l) e^{-i\omega lk}\\
        &=\frac{1}{N+1}\sum_{l=0}^N e^{-i\omega lk}\Big[\prod_{j=1}^N(1-\tau_j+\tau_j.e^{i\omega l})\Big]
        \label{pdf}
        \end{aligned}
    \end{equation}
   
\subsection{Throughput}
Let us introduce the indicator random variables $X_i, i=1,\dots,M$ such that
\begin{equation}
    X_i =
    \begin{cases}
      1 & \text{if only one out of $k$ nodes transmit in $i^{th}$ channel }\\
      0 & \text{otherwise}\\
    \end{cases}  
    \label{Xi}
    \end{equation}
    \textcolor{red}{Probability that only one node out of $k$ nodes transmitting in the $i^{th}$ channel can be determined as follows. One out of $k$ nodes will choose $i^{th}$ channel with probability $1/M$, since there are $M$ channels, and the rest $k-1$ nodes chooses other channels with probability $(1-1/M)^{k-1}$ as given below}
    \begin{equation}
\mathbb{P}(X_i=1) =\binom{k}{1} \frac{1}{M}(1-\frac{1}{M})^{k-1}
\label{P(Xi)}
\end{equation}
      $Y(k,M)$ be the random variable indicating the number of successful transmissions when $k$ nodes transmit in $M$ channels in a time slot and hence
   $ Y(k,M)= X_1+X_2+\dots+X_M $. The conditional expectation of the number of successful transmissions given $k$ nodes are transmitting in a time slot over $M$ channels is given by
\begin{equation}
\begin{aligned}
    \mathbb{E}(Y(k,M))=&\sum_{i=1}^M\mathbb{E}(X_i)=\sum_{i=1}^M1\mathbb{P}(X_i=1)+0\mathbb{P}(X_i=0)\\=&\sum_{i=1}^Mk\frac{1}{M}\Big(1-\frac{1}{M}\Big)^{k-1}=k\Big(1-\frac{1}{M}\Big)^{k-1}
    \end{aligned}
    \label{E(Y)}
\end{equation}
System throughput denoted by $T$ is the average number of successful transmissions in a slot. \textcolor{red}{$T$ is obtained from \eqref{E(Y)} by averaging $\mathbb{E}(Y(k,M))$ over the distribution of number of transmitting nodes, $P_t(k,N)$.}
\begin{equation}
\begin{aligned}
    T=&\sum_{k=0}^N P_t(k;N)\mathbb{E}(Y(k,M))
    =&\sum_{k=0}^N P_t(k;N)k\Big(1-\frac{1}{M}\Big)^{k-1}
    \end{aligned}
    \label{T}
\end{equation}
Equation \eqref{T} gives the system throughput of the considered TSCH network. 
\subsection{Delay}
In this subsection we derive the average delay experienced by a packet in the TSCH network. As the transmission probability of each node is different, the average delay experienced by different nodes are different. Let us derive the delay experienced by an arbitrary node $i$, whose transmission probability is $\tau_i$.
Let $P_t(k;N-1)$ denotes the probability of $k$ other nodes transmitting in a slot apart from the tagged node. This is calculated from \eqref{pdf} as follows
\begin{equation}
       P_t(k;N-1)=\frac{1}{N}\sum_{l=0}^{N-1} e^{-i\omega lk}\Big[\prod_{j=1,j\neq i}^N(1-\tau_j+\tau_j.e^{i\omega l})\Big]
       \label{pt}
\end{equation}
where $\omega=\frac{2\pi}{N}$. Let \textcolor{red}{$P(S/k)$} indicate the probability that a tagged transmission is successful given there are $k$ other transmissions. This is the probability that none of the $k$ transmissions choose the same channel as that of the tagged channel and is given as follows
\begin{equation}
    \textcolor{red}{P(S/k)}=(1-\frac{1}{M})^k
    \label{p(succ)}
\end{equation}

Let $p_i$ be the probability that a packet transmitted by a tagged node is successful. \textcolor{red}{Therefore $p_i$ can be calculated from \eqref{pt} and \eqref{p(succ)} by averaging $P(S/k)$ over the distribution of $k$ given by $P_t(k;N-1)$ as follows}
\begin{equation}
    \begin{aligned}
        p_i=&\sum_{k=0}^{N-1}P_t(k;N-1)(1-\frac{1}{M})^k
        \label{pi}
         \end{aligned}
        \end{equation}  
        
Let the random variable $S_i$ denotes the service time of a packet in units of slots. $P(S_i=n)$ is the probability that a packet takes $n$ slots to get served. In other words it experiences $n-1$ failure slots followed by a successful transmission.
\begin{equation}
\begin{aligned}
  P(S_i=n)=(1-\tau_i p_i)^{n-1}\tau_i p_i \qquad n=1,2,3,\dots
    \end{aligned}
    \label{Dpdf}
\end{equation}
From \eqref{Dpdf}, it can seen that the service times are geometrically distributed. The mean and second moment of the service time are calculated as below \cite{ross1996stochastic}, 
\begin{equation}
    \begin{aligned}
        \overline{S_i}=\mathbb{E}[S_i]=&\frac{1}{\tau_i p_i}
    \end{aligned}
    \label{meanS}
\end{equation}
\begin{equation}
    \overline{S_i^2}=\mathbb{E}[S_i^2]=\frac{2-\tau_i p_i}{\tau_i^2p_i^2}
    \label{secmom}
\end{equation}
Consider that each node $i$ maintains a queue to which packets arrive according to a Poisson process with rate $\lambda_i$ and packet service times having a geometric distribution. Hence the total delay experienced by a packet from the time it entered the queue till it get served can be calculated by the $Pollaczek-Khinchin$ (P-K) $formula$ for the $M/G/1$ \cite{medhi2002stochastic}. With the average service time and the second moment of service time given by \eqref{meanS} and \eqref{secmom} respectively, the total delay experienced by a packet of $i^{th}$ node is given as
\begin{equation}
    D_i=S_i+\frac{\lambda_i\overline{S_i^2}}{2(1-\lambda_i\overline{S_i})}
    \label{delay}
\end{equation}
\begin{figure}
        \centering
        \includegraphics[width=.85\linewidth]{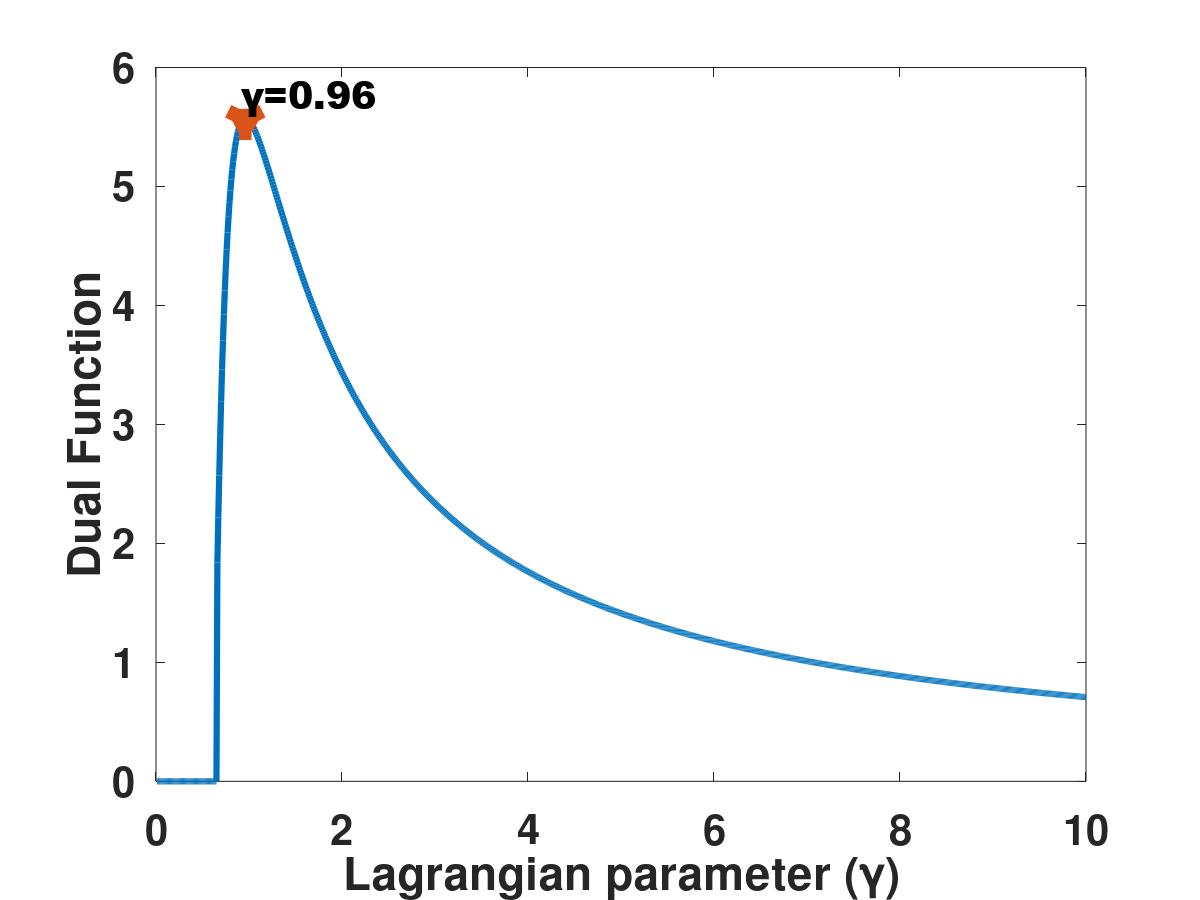}
        \caption{Dual Function $\mathcal{G}(\gamma)$ in Eq. \eqref{dual2} }
        \label{fig:dualvsgam}
        
    \end{figure}
    
    \begin{figure}
        \centering
        \includegraphics[width=.77\linewidth]{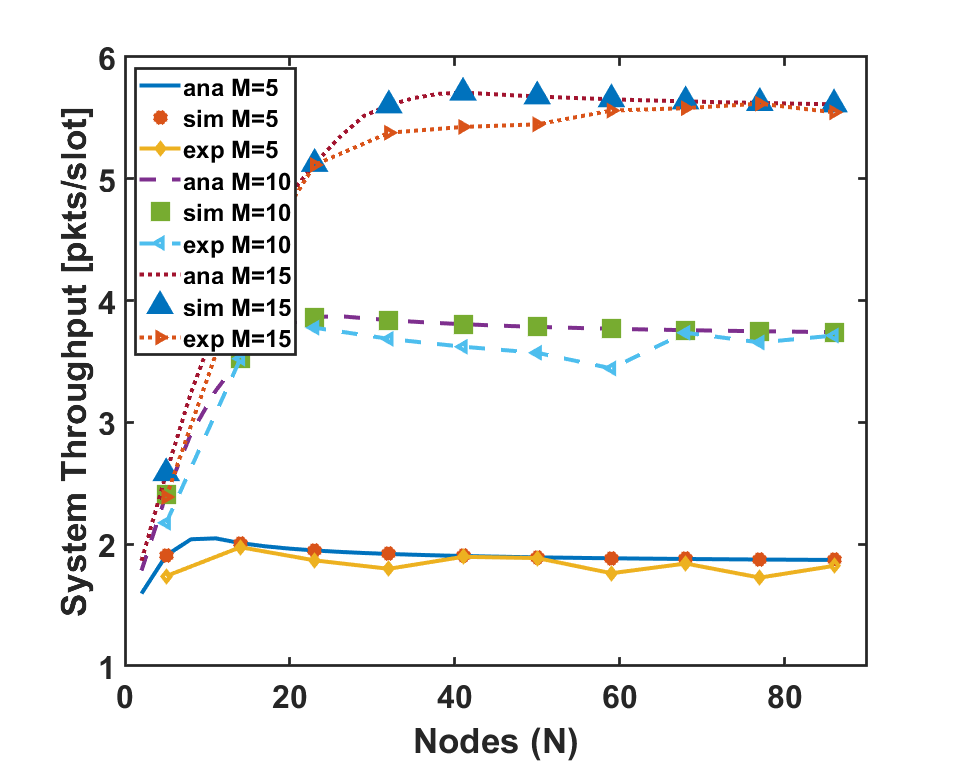}
        \caption{System Throughput (in packets per slot) against the number of nodes, $N$, for a homogeneous network}
        \label{fig:SvsN}
    \end{figure}
\subsection{Energy}
    %Let $E_{tx}$ be the energy spent in transmission of a packet by a node and is measured in terms of the number of transmission attempts until success. If the successful transmission occurs at $n^{th}$ slot, then until $n-1^{th}$ slot the packet was either differed from transmission or was collided. Let us define an indicator random variable $Y_j$, with $j$ as slot index, as follows.
    Consider that the tagged node takes $n$ slots to transmit its packet successfully, whose distribution is given in \eqref{Dpdf}. Let's define an indicator random variable $Y_j$, $j=1,2,\dots,n-1$ for all the unsuccessful slots. $Y_j$ indicates the reason for the failure of transmission made by the tagged node in $j^{th}$ slot.
  \begin{equation}
      Y_j =
    \begin{cases}
      1 & \text{if transmission in $j^{th}$ slot encounter collision}\\
      0 & \text{node differs from transmission in $j^{th}$ slot}
    \end{cases}  
    \label{Yj}
  \end{equation}
  \begin{figure}
        \centering
        \includegraphics[width=.8\linewidth]{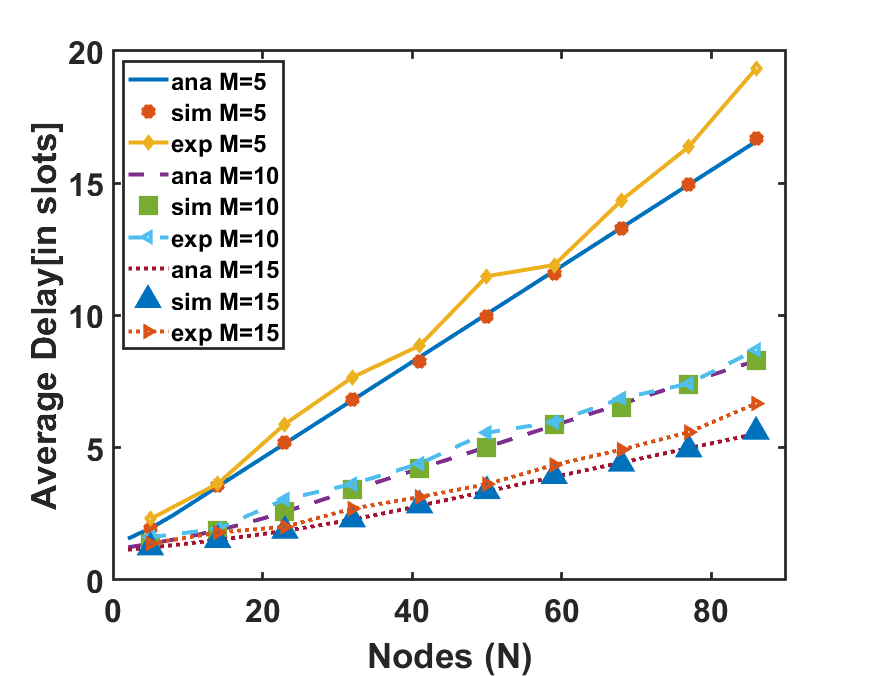}
        \caption{Average Delay against $N$ for a homogeneous network}
        \label{fig:DvsN}
        
    \end{figure}
    \begin{figure}
        \centering
        \includegraphics[width=.8\linewidth]{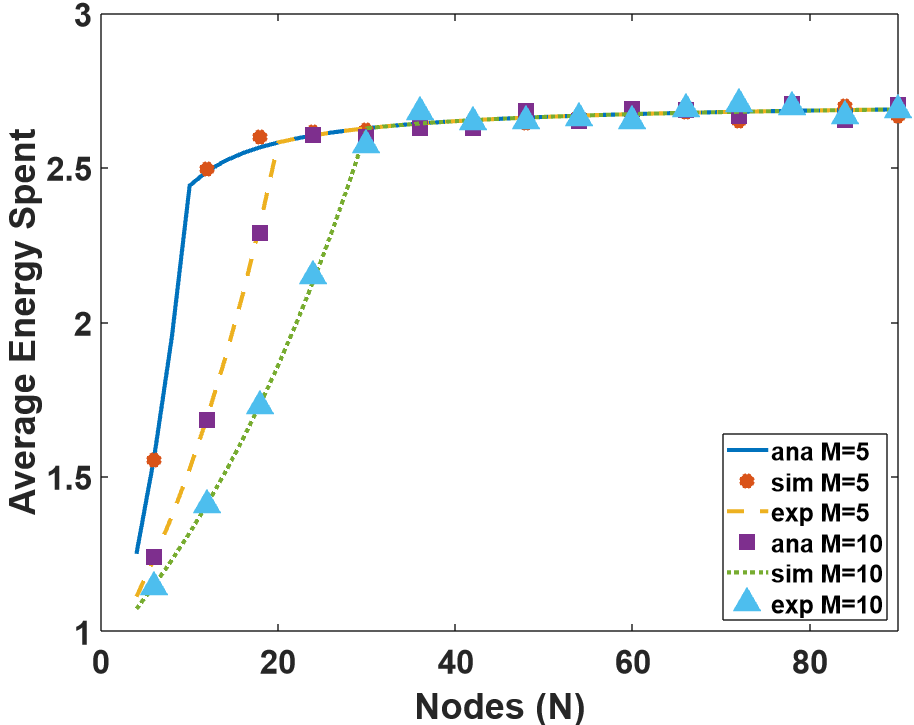}
        \caption{Average Energy spent by a node against N for a homogeneous network}
        \label{fig:EvsN}
        
    \end{figure} 
  Probability that the tagged node transmits a packet and collision occurs in $j^{th}$ slot is $P(Y_j=1)=\frac{ \tau(1-p)}{1-\tau p}$. The expected number of collisions $C$ a packet experiences before getting successfully transmitted in $n^{th}$ slot is given by 
  \begin{equation}
  \begin{aligned}
      {C}\overset{a}{=}&\mathbb{E}\Big[\sum_{j=0}^{n-1}Y_j\Big]\\
      \overset{b}{=}&\mathbb{E}[n-1]\mathbb{E}[Y_j]\\
      =&(\frac{1}{\tau p}-1)*(\frac{\tau(1-p)}{1-\tau p})\\
      =&\frac{1}{p}-1
      \end{aligned}
      \label{C}
  \end{equation}
  $\eqref{C}.b$ is obtained from $\eqref{C}.a$ by Wald's theorem\cite{ross1996stochastic}, using the fact that $Y_j$'s are IID and independent to $n$. Let $E_{tx}$ be the energy spent per transmission attempt.In total,  a node takes $1+C$ transmission attempts for successfully transmitting a packet. Hence the average energy  \(\mathcal{E}\) spent to successfully transmit the packet of tagged node is 
  \begin{equation}
      \mathcal{E}=E_{tx}*(1+C)=\frac{E_{tx}}{p}
      \label{Energy}
  \end{equation}
    
    \begin{figure}
        \centering
        \includegraphics[width=.8\linewidth]{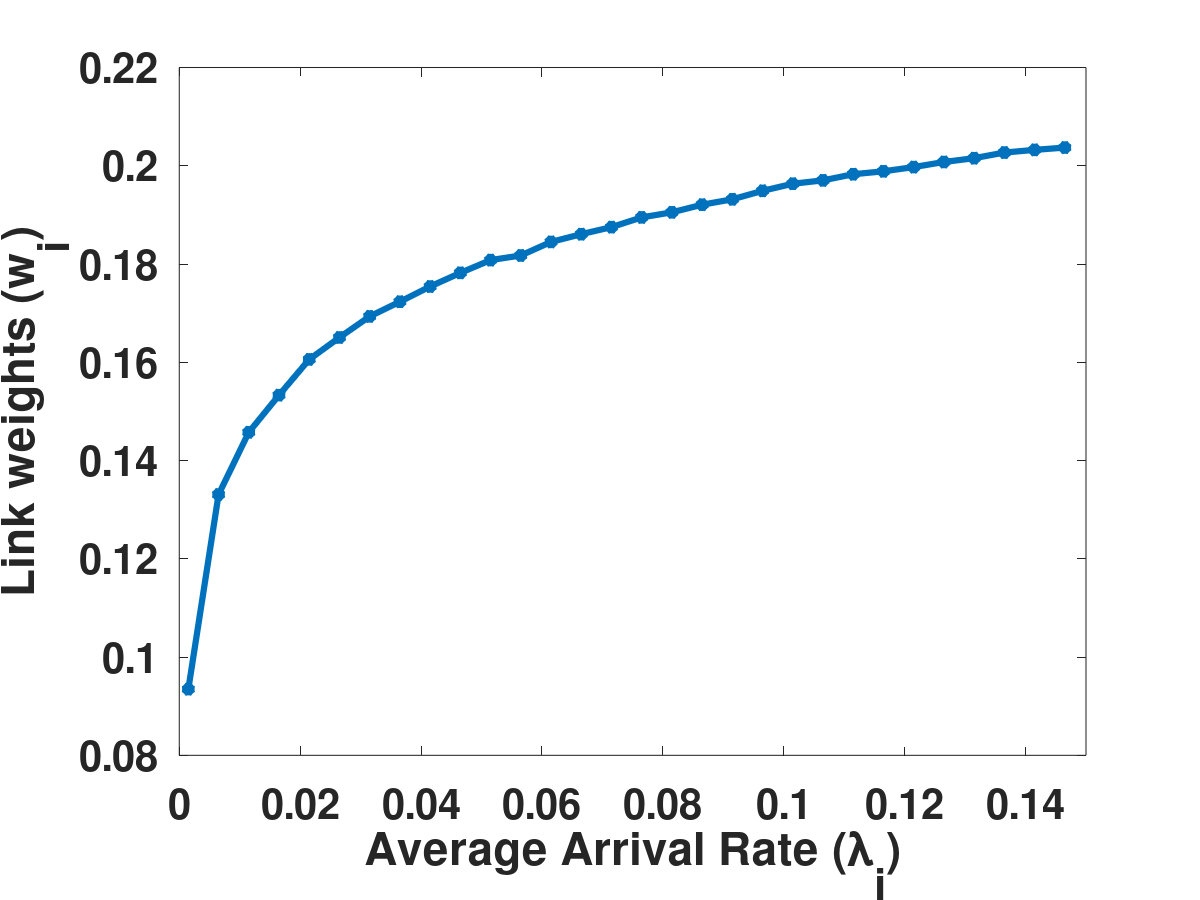}
        \caption{Normalised link weights against the average packet arrival rates in a TSCH network with $N=86$, $M=15$ }
        \label{fig:wvslam}
         \end{figure}
\section{Results and Discussion}\label{results}

\subsection{Experimental Setup and Simulation Methodology}\label{simulation}
\textcolor{teal}{We implement our CBPF transmission scheme in Contiki OS \textcolor{blue}{\cite{dunkels2004contiki}}, an open source operating system for IoT network devices. Contiki uses hybrid modular architecture based event -driven model. To verify analytical results, we evaluate the proposed CBPF scheme (Contiki implementation) on the IoT-LAB \textcolor{blue}{\cite{adjih2015fit}} public testbed installed in Grenoble, France.  We deploy our contiki implementation in 86 M3 nodes (ARM Cortex M3) whose radio chip was designed to implement the MAC layer of the IEEE 802.15.4 standard. The COAP (Constrained application protocol) generates the payload data at each node Application payload is carried in UDP/IPv6 datagrams over 6LoWPAN. Contiki-RPL is used at the routing layer to construct the network topology. Objective function of RPL is chosen as MRHOF \cite{gnawali2012minimum} which builds a tree topology using estimated link transmissions (ETX) and hop count of nodes. The application payload data is carried in UDP/IPv6 datagrams over 6LoWPAN adaption layer. 6LoWPAN protocol enables the transmission of IPv6 datagrams over the IEEE 802.15.4 radio links by adding an adaptation layer between the data link layer (IEEE 802.15.4 e TSCH) and the network layer \cite{municio2019simulating}. The CBPF  transmission scheme proposed in this paper is used at the MAC layer to trigger data transmission at nodes to the border router.}

\textcolor{teal}{CBPF uses the default slotframe structure define in the IEEE 802.15.4e standard. The CBPF control packets containing the queuelength details, weights of the nodes and the schedule are transmitted in the broadcast slots of the slotframe. The RPL control packets are transmitted based on the trickle timer mechanism as described in RFC 6206 \cite{levis2011trickle} of the standard.}

\textcolor{red}{Each Cortex M3 node uses Tx power of -16.65 dBm and Rx power of -15.98dBm.  Each node accommodate 4 sensors (light, pressure and temperature, magnetometer and gyroscope) which contribute to the traffic generated at that node. The control node (border router) records sensors data collected over the network from all nodes in TSCH network over multiple channels. Different traffic rates at different nodes required for our analysis of heterogeneous network is made possible by setting different update frequencies as well as by enabling only a subset of available sensors at a node. The sniffed data at the border router is analysed using Wireshark \textcolor{blue}{\cite{orebaugh2006wireshark}} and we calculate the necessary performance metrics of the network. Each experiment lasts for 1 hour so that sufficient number of packet transactions occur and $10\sim15$ such iterations are run over which the results are averaged.}

 \begin{figure}
        \centering
        \includegraphics[width=.8\linewidth]{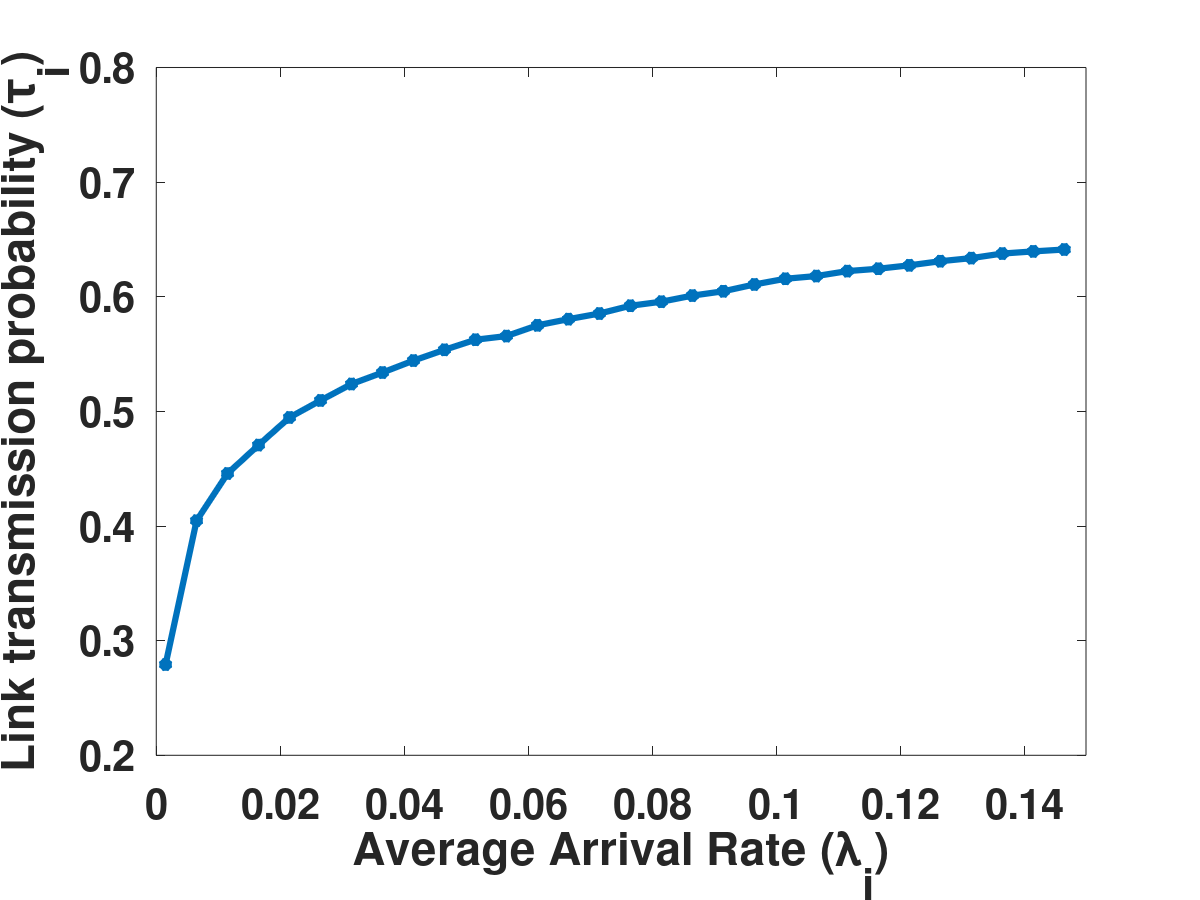}
        \caption{Optimal link transmission probability against the average packet arrival rates in a TSCH network with $N=86$, $M=15$}
        \label{fig:lamvstr}
    \end{figure}
    
\textcolor{red}{Besides, we have performed Monte-Carlo simulations of TSCH network for the scheduled and unscheduled cases using MATLAB. Network consisting of $N$ nodes ranging from 1 to 30, all transmitting data to a border router is considered. The number of channels $M$ is fixed as 5, 10 or 15. \textcolor{teal}{We consider that the nodes transmit data to border router over a maximum of 5 hops.} Throughput, delay and energy spent per transmission are obtained by averaging over 10000 iterations. Analytical results are obtained through numerical computation of the  derived equations  using MATLAB in line with the simulation parameters. The estimated link transmission is obtained through the simulations performed in 6TISCH simulator \cite{7899546}.}

\textcolor{red}{In this section, we present the simulation and experimental results for the CBPF algorithm proposed in section \ref{sec:algorithm} We compare CBPF scheme with the minimal scheduling function (MSF) \textcolor{blue}{\cite{vilajosana2017minimal}} described in the standard and also with a link scheduling algorithm, ALICE\textcolor{blue}{\cite{kim2019alice}}}.
 \subsection{Homogeneous Network}   
 \label{results:homo}
    We first consider a homogeneous network where all links are of equal importance i.e., all the links have same traffic arrival rate $\lambda_i$. Hence we consider $\lambda_{i}=0.4$ for all links to simulate a network with number of nodes $N=86$ and the number of channels $M=15$. Since all links are homogeneous the weights of all links according to \eqref{weights} are found to be same and the weight is $w_i=0.18257$. With these given weights, we solve the optimization problem to find the transmission probabilities of network that optimizes the objective function. Fig. \ref{fig:dualvsgam} shows that the dual function achieves its maximum at $\gamma=0.96$. The transmission probabilities of the nodes determined using \eqref{quadratic} for $\gamma=0.96$ is $\tau_i=0.1744$. With this solution of the transmission probabilities, the system throughput achieved is 5.6115 packets/slot.

     \begin{figure}
        \centering
        \includegraphics[width=.8\linewidth]{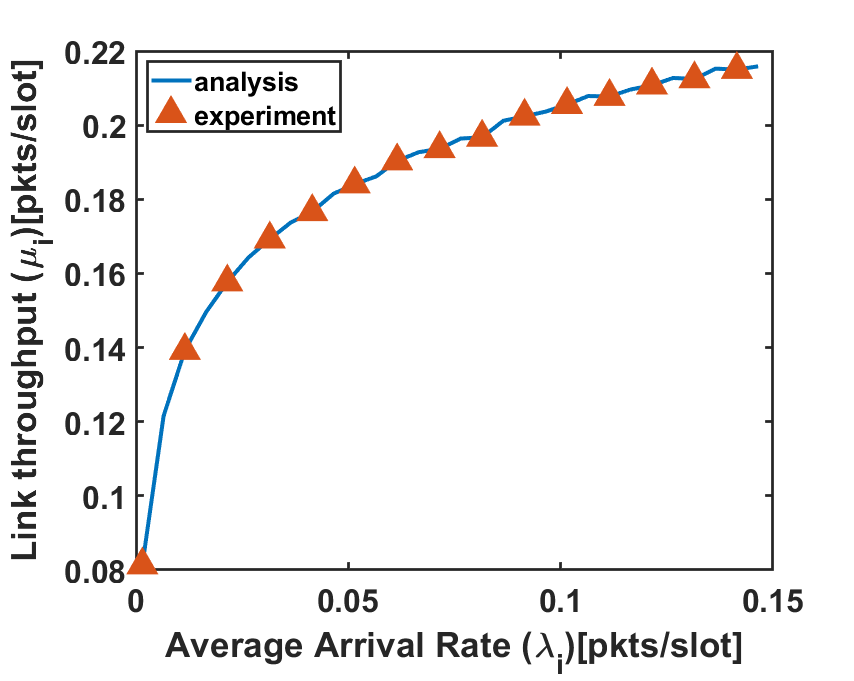}
        \caption{Optimal link throughput against the average packet arrival rates in a TSCH network with $N=86$, $M=15$}
        \label{fig:lamvsmu}
    \end{figure}
    
    In Fig. \ref{fig:SvsN}, we plot the average system throughput with variation in the network size. The system throughput increases with the network size initially until it attains a maximum and gets saturated for the further increase in network size. The maximum value and hence the saturation occurs when the number of nodes in network is same as the number of available channels i.e., $N=M$. In traditional slotted aloha networks, the throughput attains its maximum at $N=M$, however it reduces for the higher values of $M$. It can also be observed that the peak throughput achieved is $T=1.85$,$3.7$ and $5.52$ for $M=5,10$ and $15$ respectively, which is the maximum achievable theoretical throughput of slotted aloha systems $(T=0.37*M)$ \cite{9261241}.

    \begin{figure}
        \centering
        \includegraphics[width=.8\linewidth]{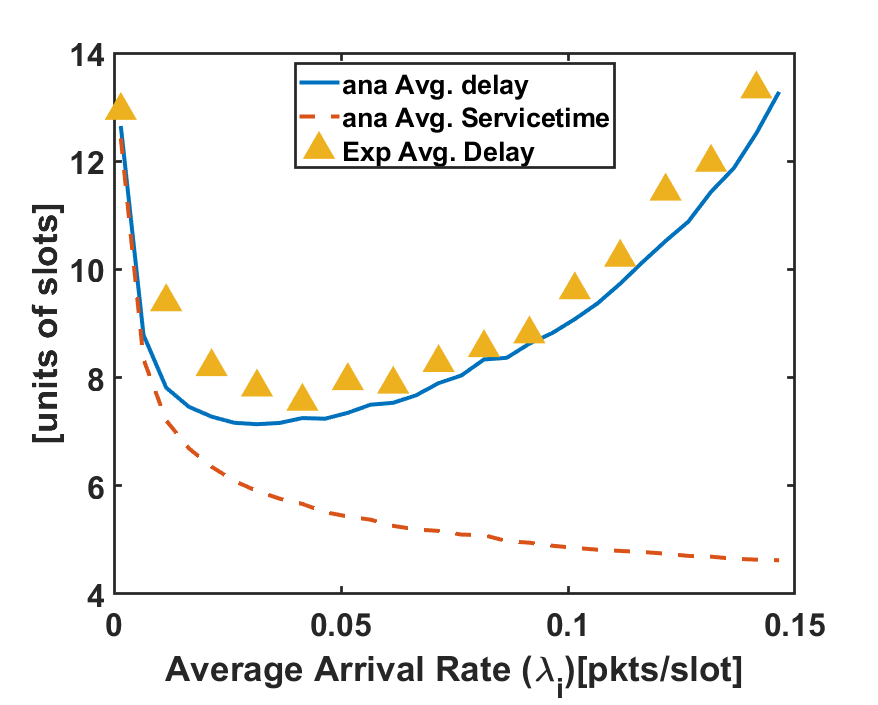}
        \caption{Service time and total delay of a packet against the arrival rate for $N=86$, $M=15$}
        \label{fig:dvslam}
        
    \end{figure}
    
    The proposed optimal transmission scheme results in a linear increase of delay with $N$ as shown in Fig. \ref{fig:DvsN}, which otherwise would be an exponential. The average number of transmissions required by a node is shown in  Fig. \ref{fig:EvsN}. The number of transmissions increases with increase in $N$ due to the increased contention but saturates after $N=M$ which can be explained as follows. From \eqref{C}, it is clear that the number of collisions experienced by a node is inversely proportional to the successful packet transmission probability $p$. Yang \textit{et al.} derived that $p=T/G$, where $T$ is the system throughput and $G$ is the traffic load i.e, the average number of transmission attempts in a slot \cite{1245995}. In our model, due to the constraint in \eqref{P2}, the load will be $G=N$ when $N\leq M$ and $G=M$ for higher $N$. Also it can be seen from Fig. \ref{fig:SvsN} that $T$ saturates after $N=M$. Putting these two facts together, the number of transmissions required also saturates after $N=M$. As the number of collisions is saturated, the energy required for a packet transmission also will be bounded and saturated. The theoretical results have been validated using simulations as well as the results obtained from the IoT-LAB testbed implementation.
 \begin{figure}
    \centering
    \includegraphics[width=.8\linewidth]{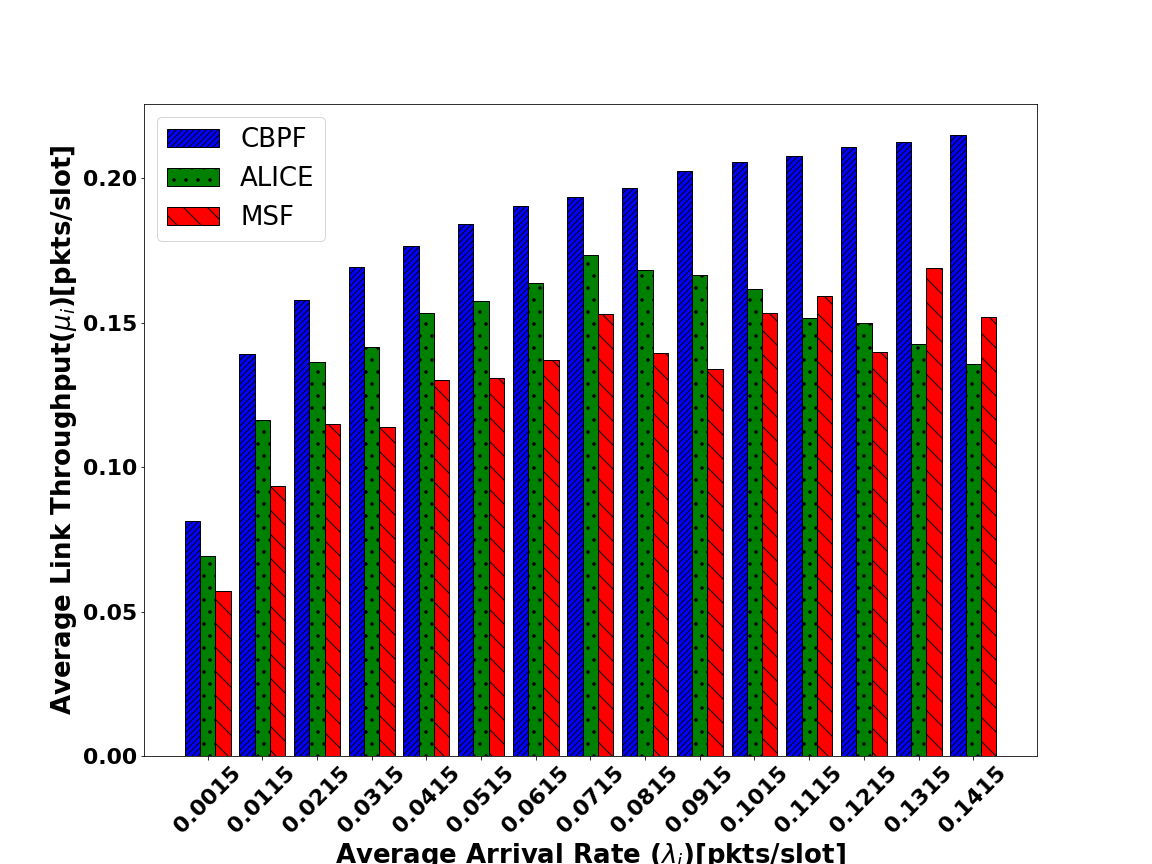}
    \caption{Throughput performance comparison of the proposed CBPF transmission scheme, ALICE and MSF for a heterogeneous network of size 86 nodes}
    \label{fig:expthcom}
\end{figure}
\begin{figure}
    \centering
    \includegraphics[width=.8\linewidth]{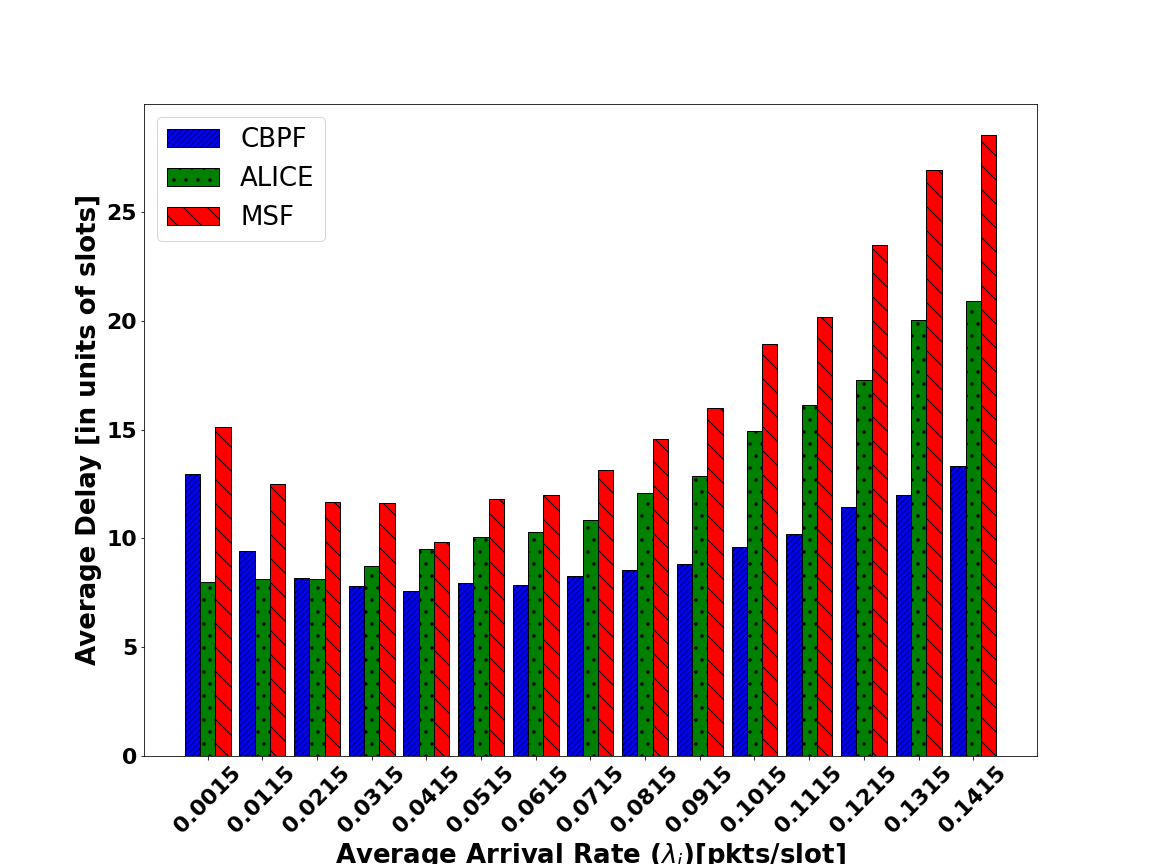}
    \caption{Delay performance comparison of the proposed CBPF transmission scheme, ALICE and MSF for a heterogeneous network of size 86 nodes}
    \label{fig:expdcom}
\end{figure}
   
\subsection{Heterogeneous Network}\label{hetresults}
\textcolor{teal}{Now, we consider a heterogeneous TSCH network where the packet arrival rates at each node in the network is specific to the application it serves. Hence the packet arrival rate for each node in the network can be different from that of the other nodes.} We consider a network consisting of $N=86$ nodes transmitting over $M=15$ channels. Each node generates packets according to a Poisson process whose mean value is different for different nodes in the range of (0, 0.2). Hence the links are assigned weights based on their queue lengths as described in Section \ref{sec:algorithm}. In Fig. \ref{fig:wvslam}, we plot the link weights against the packet arrival rate of the links. As the weightage function in \eqref{weights} is logarithmic with the queue length, the weights assigned to the links also follow a logarithmic curve with respect to $\lambda_i$ in expectation as shown in Fig. \ref{fig:wvslam}.  The proposed contention based scheduling algorithm calculates the transmission probabilities of the links as shown in Fig. \ref{fig:lamvstr} such that the weighted proportional throughput is maximized. With the obtained optimal transmission probabilities, the throughput of links is plotted in Fig. \ref{fig:lamvsmu}.

\begin{figure}
    \centering
    \includegraphics[width=.8\linewidth]{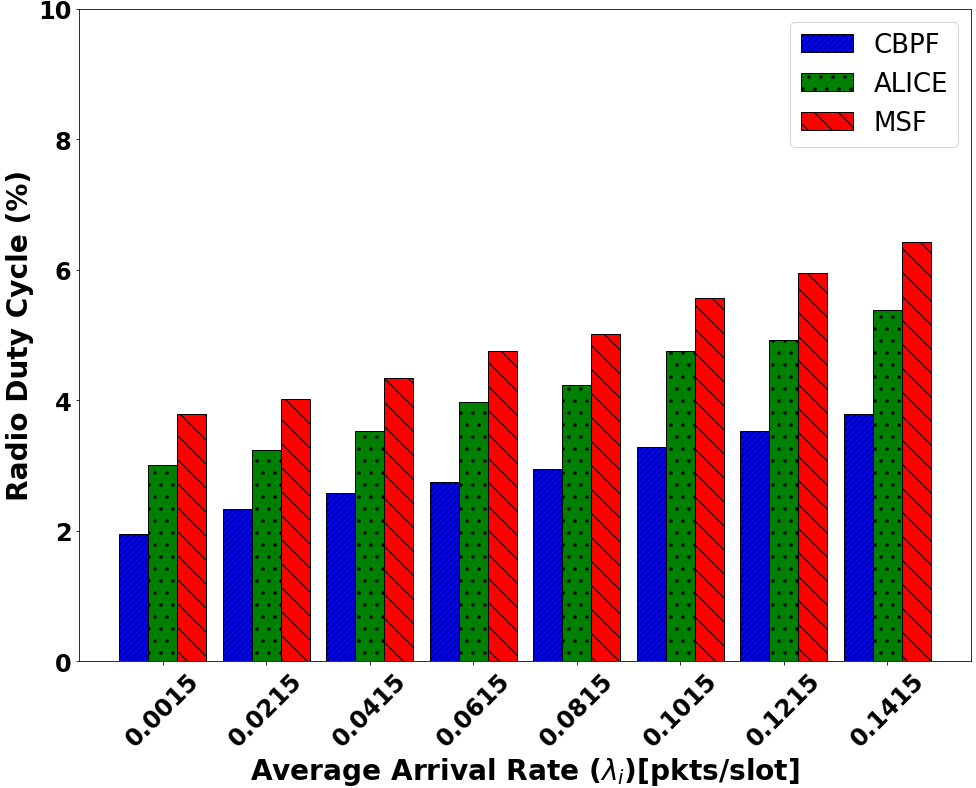}
    \caption{\textcolor{teal}{Radio duty cycle performance comparison of the proposed CBPF transmission scheme, ALICE and MSF for a heterogeneous network of size 86 nodes}}
    \label{fig:exprcom}
    \vspace{-4mm}
\end{figure}

As the objective is to maximize the weighted proportional throughput, the throughput of each link should be proportional its weight. Therefore, the link throughput should vary linearly with the weights. Hence the link throughput varies in logarithmic fashion with the packet arrival rate as shown in Fig. \ref{fig:lamvsmu}. In Fig. \ref{fig:dvslam}, we plot the service time and the total delay experienced by a packet against the arrival rate. We have validated the theoretical results with the IoT-LAB testbed results. As the arrival rate increases, the weight assigned also increases which results in the higher link throughput. Therefore the time required by a packet in the higher weight link is less compared to that of the lower weighted links. But on the other hand, as the arrival rate is higher, the number of packets in the queue builds up quickly leading to a larger waiting time of a newly arrived packet. While the links with very low arrival rate will almost see a zero queuing delay since the packets are serviced as soon as they arrive in the queue. Total delay experienced by a packet is the sum of its queuing delay and the service time. Since queuing delay is very less at the nodes with low arrival rates, service time ans the total delay coincides with each other.      

\subsection{Comparative Analysis}
\textcolor{red}{In this subsection, we compare the performance of our CBPF transmission scheme, presented in \ref{hetresults}, with the minimal scheduling function (MSF) \cite{vilajosana2017minimal} and ALICE \cite{kim2019alice}. Throughput and delay performance of the three algorithms are presented in Fig. \ref{fig:expthcom} and Fig. \ref{fig:expdcom} respectively. These results are collected from the IoT-LAB test bed by implementing MSF and ALICE scheduler in the deployed protocol stack. MSF allocates cells to links without taking into consideration of the possible collisions in network. Beyond that nodes have to poll for their share of resources following the 6P transactions, which causes additional collisions. Hence our algorithm outperforms MSF in all aspects, but following the same trend.  But in case of ALICE, the transmission conflicts are taken into consideration while designing the schedule. However due to the autonomous nature of ALICE, a node doesn't have the knowledge of neighbouring nodes, resulting in the sub-optimal schedule. Besides, ALICE does a mapping of cells to be used by links for transmission. Attaining an optimal mapping between links and the cells in slotframe is a NP hard problem. Hence, the schedule becomes highly inefficient for large traffics and large network sizes causing larger latency as can be observed from Fig. \ref{fig:expdcom}.} \textcolor{teal}{In Fig. \ref{fig:exprcom}, we present the radio duty cycle performance of CBPF in comparison with MSF and ALICE. Since the proposed CBPF gives the optimal frequency of transmissions, the nodes in the network keep their radio on for the optimal number of slots and hence the duty cycle is kept low in comparison to MSF and ALICE, which are sub-optimal schedules. Hence, optimal transmission scheme ensures energy efficient communication.}
   
 \section{Conclusion}
 \label{conclusion}
 The future of industrial communication is relied on wireless networks and Internet of Things. As the number of devices connected to the network is increasing rapidly and the they generate sporadic traffic, reservation based scheduling is highly inefficient. Hence we propose a contention based transmission scheme for the TSCH network such that the weighted proportional throughput of the network is maximised. Also the proposed transmission scheme limits the number of collisions experienced by a packet, the energy consumption of the nodes during communication can be reduced enormously. We have implemented our proposed CBPF transmission scheme on a real time test bed IoT-Lab and have compared the performance of our algorithm with the existing schemes namely ALICE and MSF. The performance results showed that the proposed transmission scheme achieves the theoretical maximum throughput even when the network size is large.

    \appendices
\section{Proof of the Convexity of weighted sum throughput function}
\label{appendix}
    \textit{Lemma 1:} The weighted sum throughput function $F=\sum_{(n,m)\in \mathcal{S}}w_{nm}log(\mu_{nm}(\boldsymbol{\tau}))$ is a concave function with respect to $\boldsymbol{\tau}$.\\
    \textit{Proof:} Let the links in the TSCH network are indexed with $i, i=1,2,\dots,S$ rather than with the source and destination nodes for simplicity. Let us denote the log rate function $log(\mu_i(\boldsymbol{\tau}))$ by $R_i(\boldsymbol{\tau})$ for link $i$. Now, the weighted sum throughput function $F$ is given in terms of $R_i(\boldsymbol{\tau})$ as follows.
    \begin{equation}
        F=\sum_{i=1}^S\mu_i R_i(\boldsymbol{\tau})
        \label{F,R_i}
    \end{equation}
    where from \eqref{mu}, $R_i(\boldsymbol{\tau})$ can be written as
    \begin{equation}
    \begin{aligned}
        R_i(\boldsymbol{\tau})&=log\Big(\tau_i \prod_{j\in\mathcal{I}_i^p}(1-\tau_j) \prod_{k\in \mathcal{I}_k^s\setminus\mathcal{I}_k^p}(1-\frac{\tau_k}{M})\Big)\\
        &=log(\tau_i)+\sum_{j\in\mathcal{I}_i^p}log(1-\tau_j)+\sum_{k\in \mathcal{I}_k^s\setminus\mathcal{I}_k^p}log(1-\frac{\tau_k}{M})
        \label{R_i}
        \end{aligned}
    \end{equation}
    Denote the Hessian of $R_i(\boldsymbol{\tau})$ as
    \begin{equation}
        \bigtriangledown^2(R_i(\boldsymbol{\tau}))=\big[a_{xy}^{(i)}\big], 1\leq x,y \leq S
        \label{hessian}
    \end{equation}
    where $a_{xy}^{(i)}$ denotes the element of $\bigtriangledown^2(R_i(\boldsymbol{\tau}))$ at the $x^{th}$ row and $y^{th}$ column and can be expressed as
    \begin{equation}
       a_{xy}^{(i)} = \frac{\partial^2(R_i(\boldsymbol{\tau}))}{\partial \tau_x \partial \tau_y}
       \label{hess}
    \end{equation}
    From \eqref{R_i} and \eqref{hess}, the elements of hessian matrix $\bigtriangledown^2(R_i(\boldsymbol{\tau}))$ are determined as,
   \begin{equation}
    a_{xy}^{(i)} =
    \begin{cases}
      -\frac{1}{\tau_i^2}& x=y=i\\
      -\frac{1}{(1-\tau_j)^2}&x=y=j, j \in \mathcal{I}_i^p\\
      -\frac{1}{M^2}\frac{1}{(1-\frac{\tau_j}{M})^2}&x=y=j, j \in \mathcal{I}_i^s\setminus\mathcal{I}_i^p\\
      0&\text{otherwise}
      \label{hessian coeff}
    \end{cases}  
    \end{equation}
    From \eqref{hessian coeff}, it is clear that all entries of the Hessian matrix are non positive. Therefore the Hessian of log rate function of a link is negative semidefinite matrix, $\bigtriangledown^2(R_i(\boldsymbol{\tau}))\preceq 0$. Hence $R_i(\boldsymbol{\tau})$ is a concave function of $\boldsymbol{\tau}$. 
    
    The function $F$, as can be seen from \eqref{F,R_i}, is the non negative weighted sum of $R_i(\boldsymbol{\tau})$. Following the well established fact that non negative weighted sum of concave functions is a concave function \cite{boyd2004convex}, it can be seen that $F$ is a concave function of $\boldsymbol{\tau}$.
   \bibliographystyle{IEEEtran}
\bibliography{references}

\begin{thebibliography}{10}
\providecommand{\url}[1]{#1}
\csname url@samestyle\endcsname
\providecommand{\newblock}{\relax}
\providecommand{\bibinfo}[2]{#2}
\providecommand{\BIBentrySTDinterwordspacing}{\spaceskip=0pt\relax}
\providecommand{\BIBentryALTinterwordstretchfactor}{4}
\providecommand{\BIBentryALTinterwordspacing}{\spaceskip=\fontdimen2\font plus
\BIBentryALTinterwordstretchfactor\fontdimen3\font minus
  \fontdimen4\font\relax}
\providecommand{\BIBforeignlanguage}[2]{{%
\expandafter\ifx\csname l@#1\endcsname\relax
\typeout{** WARNING: IEEEtran.bst: No hyphenation pattern has been}%
\typeout{** loaded for the language `#1'. Using the pattern for}%
\typeout{** the default language instead.}%
\else
\language=\csname l@#1\endcsname
\fi
#2}}
\providecommand{\BIBdecl}{\relax}
\BIBdecl

\bibitem{hashemi2014intra}
M.~Hashemi, W.~Si, M.~Laifenfeld, D.~Starobinski, and A.~Trachtenberg,
  ``Intra-car multihop wireless sensor networking: a case study,'' \emph{IEEE
  Communications Magazine}, vol.~52, no.~12, pp. 183--191, 2014.

\bibitem{elsts2020empirical}
A.~Elsts, S.~Kim, H.-S. Kim, and C.~Kim, ``An empirical survey of autonomous
  scheduling methods for tsch,'' \emph{IEEE Access}, vol.~8, pp.
  67\,147--67\,165, 2020.

\bibitem{Guglielmo2016IEEE8}
D.~D. Guglielmo, S.~Brienza, and G.~Anastasi, ``Ieee 802 . 15 . 4 e : a
  survey,'' 2016.

\bibitem{thubert2015architecture}
P.~Thubert, ``An architecture for ipv6 over the tsch mode of ieee 802.15. 4,''
  \emph{Internet-Draft draft-ietf-6tisch-architecture-28(work in progress)},
  october 2019.

\bibitem{winter2012rpl}
T.~Winter, P.~Thubert, A.~Brandt, J.~Hui, R.~Kelsey, P.~Levis, K.~Pister,
  R.~Struik, J.~Vasseur, R.~Alexander \emph{et~al.}, ``Rpl: Ipv6 routing
  protocol for low-power and lossy networks. rfc 6550,'' 2012.

\bibitem{9203871}
Y.~{Ha} and S.~H. {Chung}, ``Enhanced 6p transaction methods for industrial
  6tisch wireless networks,'' \emph{IEEE Access}, vol.~8, pp.
  174\,115--174\,131, 2020.

\bibitem{kharb2019survey}
S.~Kharb and A.~Singhrova, ``A survey on network formation and scheduling
  algorithms for time slotted channel hopping in industrial networks,''
  \emph{Journal of Network and Computer Applications}, vol. 126, pp. 59--87,
  2019.

\bibitem{8481452}
M.~O. {Ojo}, S.~{Giordano}, D.~{Adami}, and M.~{Pagano}, ``Throughput
  maximizing and fair scheduling algorithms in industrial internet of things
  networks,'' \emph{IEEE Transactions on Industrial Informatics}, vol.~15,
  no.~6, pp. 3400--3410, 2019.

\bibitem{adjih2015fit}
C.~Adjih, E.~Baccelli, E.~Fleury, G.~Harter, N.~Mitton, T.~Noel,
  R.~Pissard-Gibollet, F.~Saint-Marcel, G.~Schreiner, J.~Vandaele
  \emph{et~al.}, ``Fit iot-lab: A large scale open experimental iot testbed,''
  in \emph{2015 IEEE 2nd World Forum on Internet of Things (WF-IoT)}.\hskip 1em
  plus 0.5em minus 0.4em\relax IEEE, 2015, pp. 459--464.

\bibitem{wang20186tisch}
Q.~Wang, X.~Vilajosana, and T.~Watteyne, ``6tisch operation sublayer (6top)
  protocol (6p),'' in \emph{Internet Requests for Comments, RFC Editor, RFC
  8480}, 2018.

\bibitem{palattella2015fly}
M.~R. Palattella, T.~Watteyne, Q.~Wang, K.~Muraoka, N.~Accettura, D.~Dujovne,
  L.~A. Grieco, and T.~Engel, ``On-the-fly bandwidth reservation for 6tisch
  wireless industrial networks,'' \emph{IEEE Sensors Journal}, vol.~16, no.~2,
  pp. 550--560, 2015.

\bibitem{vergados2017toward}
D.~J. Vergados, N.~Amelina, Y.~Jiang, K.~Kralevska, and O.~Granichin, ``Toward
  optimal distributed node scheduling in a multihop wireless network through
  local voting,'' \emph{IEEE Transactions on Wireless Communications}, vol.~17,
  no.~1, pp. 400--414, 2017.

\bibitem{8576990}
H.~{Park}, H.~{Kim}, K.~T. {Kim}, S.~{Kim}, and P.~{Mah}, ``Frame-type-aware
  static time slotted channel hopping scheduling scheme for large-scale smart
  metering networks,'' \emph{IEEE Access}, vol.~7, pp. 2200--2209, 2019.

\bibitem{kim2019alice}
S.~Kim, H.-S. Kim, and C.~Kim, ``Alice: Autonomous link-based cell scheduling
  for tsch,'' in \emph{Proceedings of the 18th International Conference on
  Information Processing in Sensor Networks}, 2019, pp. 121--132.

\bibitem{TESLA}
{Jeong, Seungbeom and Paek, Jeongyeup and Kim, Hyung-Sin and Bahk, Saewoong},
  ``{TESLA: Traffic-Aware Elastic Slotframe Adjustment in TSCH Networks},''
  \emph{{IEEE Access}}, vol.~7, pp. 130\,468--130\,483, 2019.

\bibitem{jeong2020ost}
{Jeong, Seungbeom and Kim, Hyung-Sin and Paek, Jeongyeup and Bahk, Saewoong},
  ``{OST: On-demand TSCH scheduling with traffic-awareness},'' in \emph{{IEEE
  INFOCOM 2020-IEEE Conference on Computer Communications}}.\hskip 1em plus
  0.5em minus 0.4em\relax IEEE, 2020, pp. 69--78.

\bibitem{CMAB}
N.~{Taheri Javan}, M.~{Sabaei}, and V.~{Hakami}, ``Ieee 802.15.4.e tsch-based
  scheduling for throughput optimization: A combinatorial multi-armed bandit
  approach,'' \emph{IEEE Sensors Journal}, vol.~20, no.~1, pp. 525--537, 2020.

\bibitem{9096372}
V.~Kotsiou, G.~Z. Papadopoulos, P.~Chatzimisios, and F.~Theoleyre, ``Ldsf:
  Low-latency distributed scheduling function for industrial internet of
  things,'' \emph{IEEE Internet of Things Journal}, vol.~7, no.~9, pp.
  8688--8699, 2020.

\bibitem{kelly1997charging}
F.~Kelly, ``Charging and rate control for elastic traffic,'' \emph{European
  transactions on Telecommunications}, vol.~8, no.~1, pp. 33--37, 1997.

\bibitem{massoulie2007structural}
L.~Massouli{\'e} \emph{et~al.}, ``Structural properties of proportional
  fairness: Stability and insensitivity,'' \emph{The Annals of Applied
  Probability}, vol.~17, no.~3, pp. 809--839, 2007.

\bibitem{7899546}
J.~D.~R. {Nepomuceno} and N.~M.~C. {Tiglao}, ``Performance evaluation of 6tisch
  for resilient data transport in wireless sensor networks,'' in \emph{Intl.
  Conference on Information Networking (ICOIN)}, Jan 2017, pp. 552--557.

\bibitem{8418137}
H.~{Ines}, ``Performance of ieee802.15.4e tsch protocol for multi-hop wireless
  sensor networks,'' in \emph{Intl. Conference on Advanced Information
  Networking and Applications Workshops}, May 2018, pp. 603--608.

\bibitem{alderisi2015simulative}
G.~Alderisi, G.~Patti, O.~Mirabella, and L.~L. Bello, ``Simulative assessments
  of the ieee 802.15. 4e dsme and tsch in realistic process automation
  scenarios,'' in \emph{IEEE 13th Intl. Conference on Industrial Informatics
  (INDIN)}.\hskip 1em plus 0.5em minus 0.4em\relax IEEE, 2015, pp. 948--955.

\bibitem{9023554}
T.~{Chen}, J.~{Diakonikolas}, J.~{Ghaderi}, and G.~{Zussman}, ``Hybrid
  scheduling in heterogeneous half- and full-duplex wireless networks,''
  \emph{IEEE/ACM Transactions on Networking}, vol.~28, no.~2, pp. 764--777,
  2020.

\bibitem{boyd2004convex}
S.~Boyd, S.~P. Boyd, and L.~Vandenberghe, \emph{Convex optimization}.\hskip 1em
  plus 0.5em minus 0.4em\relax Cambridge university press, 2004.

\bibitem{ross1996stochastic}
S.~M. Ross, ``Stochastic processes 2nd edition john wiley \& sons,'' 1996.

\bibitem{medhi2002stochastic}
J.~Medhi, \emph{Stochastic models in queueing theory}.\hskip 1em plus 0.5em
  minus 0.4em\relax Elsevier, 2002.

\bibitem{dunkels2004contiki}
A.~Dunkels, B.~Gronvall, and T.~Voigt, ``Contiki-a lightweight and flexible
  operating system for tiny networked sensors,'' in \emph{29th annual IEEE
  international conference on local computer networks}.\hskip 1em plus 0.5em
  minus 0.4em\relax IEEE, 2004, pp. 455--462.

\bibitem{gnawali2012minimum}
O.~Gnawali and P.~Levis, ``The minimum rank with hysteresis objective
  function,'' \emph{RFC 6719}, 2012.

\bibitem{municio2019simulating}
{Municio, Esteban and Daneels, Glenn and Vu{\v{c}}ini{\'c}, Mali{\v{s}}a and
  Latr{\'e}, Steven and Famaey, Jeroen and Tanaka, Yasuyuki and Brun, Keoma and
  Muraoka, Kazushi and Vilajosana, Xavier and Watteyne, Thomas}, ``{Simulating
  6TiSCH networks},'' \emph{{Transactions on Emerging Telecommunications
  Technologies}}, vol.~30, no.~3, p. e3494, 2019.

\bibitem{levis2011trickle}
{Levis, Philip and Clausen, Thomas and Hui, Jonathan and Gnawali, Omprakash and
  Ko, JeongGil}, ``{The trickle algorithm},'' \emph{{Internet Engineering Task
  Force, RFC6206}}, 2011.

\bibitem{orebaugh2006wireshark}
A.~Orebaugh, G.~Ramirez, and J.~Beale, \emph{Wireshark \& Ethereal network
  protocol analyzer toolkit}.\hskip 1em plus 0.5em minus 0.4em\relax Elsevier,
  2006.

\bibitem{vilajosana2017minimal}
X.~Vilajosana, K.~Pister, and T.~Watteyne, ``Minimal ipv6 over the tsch mode of
  ieee 802.15. 4e (6tisch) configuration,'' 2017.

\bibitem{9261241}
J.~{Choi}, ``On improving throughput of multichannel aloha using preamble-based
  exploration,'' \emph{Journal of Communications and Networks}, vol.~22, no.~5,
  pp. 380--389, 2020.

\bibitem{1245995}
{Yang Yang} and T.~.~P. {Yum}, ``Delay distributions of slotted aloha and
  csma,'' \emph{IEEE Transactions on Communications}, vol.~51, no.~11, pp.
  1846--1857, 2003.

\end{thebibliography}

\end{document}